\documentclass[12pt]{article}
\usepackage{graphicx}
\newenvironment{sciabstract}{%
\begin{quote} 
}
{\end{quote}}

\newcounter{lastnote}

\newcommand{\non}{\nonumber \\}

\newcommand{\alp}{\alpha}     
\newcommand{\gam}{\gamma}     \newcommand{\del}{\delta}

\newcommand{\kap}{\kappa}     
   \newcommand{\sig}{\sigma}
 
   \newcommand{\ome}{\omega}

\newcommand{\Gam}{\Gamma}     
     
\newcommand{\Sig}{\Sigma}

\newcommand{\cA}{{\cal A}}

    \newcommand{\cN}{{\cal N}}
\newcommand{\cO}{{\cal O}}    
    
    \newcommand{\cT}{{\cal T}}

\newcommand{\pa}{\partial}

\renewcommand{\dag}{^{\dagger}}

\newcommand{\rar}{\rightarrow}

\newcommand{\one}{1\!\!1}
\newcommand{\hlf}{\frac{1}{2}}

\title{\bf \Large String Theory, Quantum Phase Transitions and the Emergent Fermi-Liquid.}

\author{Mihailo \v{C}ubrovi\'{c}${}^1$,  Jan Zaanen${}^1$, Koenraad Schalm${}^{1\ast}$
\\
\\
\normalsize{\em $^{1}$Institute-Lorentz for Theoretical Physics}\\
\normalsize{\em Leiden University}\\
\normalsize{\em PO Box 9506, Leiden, The Netherlands}\\
\\
\normalsize{$^\ast$
E-mail: {\tt cubrovic, jan, kschalm@lorentz.leidenuniv.nl}}
}

\date{}

\begin{document} 


\baselineskip18pt
%\renewcommand{\baselinestretch}{1.8}


\maketitle


\begin{sciabstract}
A central 
problem in quantum condensed matter physics is the critical theory governing the zero temperature quantum phase transition between strongly
renormalized Fermi-liquids as found in heavy fermion intermetallics and possibly high Tc superconductors.  
We present here results showing 
that the mathematics of string theory is capable of describing such 
fermionic quantum critical states.  Using the Anti-de-Sitter/Conformal Field Theory (AdS/CFT)
correspondence to relate fermionic quantum critical fields to a gravitational problem, we compute 
the spectral functions of fermions in the field theory. By increasing the fermion density away from the relativistic
quantum critical point, a state emerges with all the features of the Fermi-liquid. 
\end{sciabstract}

\newpage
\noindent
Quantum many-particle-physics lacks a general mathematical theory to deal with fermions at finite density.
This is known as the  ``fermion-sign-problem'': 
there is no recourse to brute force lattice models as 
the statistical path integral methods  
that work for any bosonic quantum field theory do not work for finite density fermi-systems. The non-probabilistic fermion problem is known
to be of exponential complexity \cite{Troyer:2004ge} and in the absence of a general mathematical 
framework all that remains is phenomenological guesswork in the form of the Fermi-liquid theory describing the state of electrons 
in normal metals and the mean-field theories 
%of the BCS variety 
describing superconductivity and other manifestations of spontaneous 
symmetry breaking.  This problem has become particularly manifest in quantum condensed matter physics with the discovery of electron
systems undergoing quantum phase transitions that are reminiscent of the bosonic quantum critical systems \cite{sachdev} but are
 governed by fermion statistics. Empirically well documented examples are found in the 'heavy fermion' intermetallics  where the zero 
 temperature transition occurs between different Fermi-liquids with quasiparticle masses that diverge at the quantum critical point \cite{JZScience}.
% It is a popular  belief that 
Such fermionic quantum critical states are believed to have a direct bearing on the problem of high $T_c$ superconductivity 
 because of the observation of quantum critical features in the normal state  of optimally doped cuprate high $T_c$ superconductors \cite{vandeMarel:2003wn,JZNature}.

A large part of the ``fermion-sign-problem'' is to understand this strongly coupled fermionic quantum critical state.
%  and such strongl For fermionic excitations, however, we have little understanding. The 
% problem in discussing any emergence of the Fermi-liquid lies in the fact that the conformal ﬁeld 
% theory at the critical point must be a strongly coupled one. Weakly interacting fermions are 
% not conformal: this is the Fermi liquid! And free fermions have no marginal operators in more 
% than one dimension. 
The emergent scale invariance and conformal symmetry at critical points is a benefit in isolating deep questions of principle. The question is how does the system get rid off the scales of Fermi-energy 
and Fermi-momentum that are intrinsically rooted in the workings of Fermi-Dirac statistics \cite{Senthil,Krueger}? 
Vice versa, how to construct a renormalization group with a relevant 'operator' that describes the emergence of a statistics controlled 
(heavy) Fermi liquid from the critical state \cite{JZScience}, or perhaps the emergence of a high $T_c$ 
%BCS-like 
superconductor?  We 
will show that a mathematical method developed  in  string theory has the capacity to answer at least some of these questions.

\paragraph*{String theory for condensed matter}
We refer to the AdS/CFT correspondence: a duality relation between classical gravitational physics in 
a $d+1$ dimensional 'bulk'  space-time with an Anti-de-Sitter (AdS) geometry 
and a strongly coupled conformal (quantum critical) field theory (CFT)
 with a large number of degrees of freedom that {occupies} a flat or spherical $d$ dimensional 'boundary' space-time. Applications of AdS/CFT 
 to quantum critical systems have already been studied in the context of the quark-gluon plasma \cite{Son:2007vk,Gubser:2009md},
 superconductor-insulator transitions \cite{Herzog:2007ij,Hartnoll:2007ih,Gubser:2008px,Hartnoll:2008vx,Hartnoll:2008kx} and cold atom systems at the feshbach resonance \cite{Son:2008ye,Balasubramanian:2008dm,Adams:2008wt} but so far the focus has been on bosonic currents (see \cite{HartnollScience,Hartnoll:2009sz} and references therein). Although AdS/CFT is convenient, in principle the groundstate or any response of a bosonic statistical field theory can also be computed
 directly by averaging on a lattice. For fermions statistical averaging is not possible because of the sign-problem. 
There are, however, indications that AdS/CFT should be able to capture finite density fermi systems as well. Ensembles described through AdS/CFT can exhibit a specific heat that scales linear with the temperature characteristic of Fermi systems \cite{Kulaxizi:2008jx}, zero sound \cite{Kulaxizi:2008jx,Karch:2008fa,Kulaxizi:2008kv} and a minimum energy for fermionic excitations \cite{Rozali:2008jx,Shieh:2008nf}.

To address the question whether AdS/CFT can describe finite density fermi-systems and the Fermi liquid in particular, we 
compute the single charged fermion propagators and the associated
 spectral functions that are measured experimentally by angular resolved photoemission (``AdS-to-ARPES'') and indirectly by scanning tunneling microscopy.
The spectral functions contain the crucial
 information regarding the nature of the fermion states. These are computed on the AdS side by solving for the on-shell
 (classical) Dirac equation in the curved AdS space-time background with sources at the boundary.  A temperature $T$ and finite 
$U(1)$ chemical potential $\mu_0$ for electric charge is imposed in the field theory
by studying the Dirac equation in the background of an AdS Reissner-Nordstrom black hole. We do so expecting that the $U(1)$ chemical potential induces a finite density of the charged fermions.  The procedure to compute the retarded CFT propagator from the dual AdS description is then well established \cite{Son:2007vk,Hartnoll:2009sz}. Compared to
the algorithm for computing bosonic responses, the treatment of Dirac waves in AdS is more delicate, but straightforward; details are provided in the 
supporting material \cite{supp}.
%supporting material. 
The equations obtained this way are solved numerically and
 the output is the retarded single fermion
propagator  $G_R(\ome,k)$ at finite $T$. Its imaginary part is the single fermion spectral function $A(\ome,k)=-\frac{1}{\pi}{\rm Im}{\rm Tr}(\, i\gamma^0G_R(\ome,k))$ that can be directly compared with ARPES experiments \cite{footnote1}.

The reference point for this comparison is the quantum critical point described by a zero chemical potential ($\mu_0 =0$), 
 zero temperature ($T=0$), conformal and Lorentz invariant field theory. Here the fermion propagators $\langle\bar{\Psi}\Psi\rangle \equiv G(\ome,k)$ are completely fixed by symmetry to be of the form (we use relativistic notation where $c=1$)
\begin{equation}
\label{eq:20}
%\hfill
G^{CFT}_{\Delta_{\Psi}}(\ome,k) \sim \frac{1}{(\sqrt{- \omega^2 + k^2})^{d-2\Delta_{\Psi}}}
% (1)
\end{equation}
with $\Delta_{\Psi}$ the scaling dimension of the fermion field. Through the AdS${}_{d+1}$/CFT${}_d$ dictionary $\Delta_{\Psi}$ is related to the mass parameter in the $d+1$-dimensional AdS Dirac equation. Unitarity bounds this mass from below in units of the AdS radius $mL=\Delta_{\Psi}-d/2\,>\,-1/2$ (we set $L=1$ in the remainder).  
The choice of which value to use for $m$ will prove essential to show the emergence of the Fermi liquid. The lower end of the unitarity bound $m=-1/2+\delta$, $\delta \ll 1$,
corresponds to introducing a fermionic conformal operator with weight  $\Delta_{\Psi}=(d-1)/2+\delta$. This equals the scaling dimension of a nearly free fermion. Despite the fact that the underlying CFT is strongly coupled, the absence of a large anomalous dimension for a fermion with mass $m=-1/2+\delta$ argues that such an operator fulfills a spectator-role and is only weakly coupled to this CFT. We will therefore use such values in our calculations. Our expectation is that the Fermi liquid, as a
system with well-defined quasiparticle excitations, can be
described in terms of weakly interacting long-range fields. 
%This reasoning should cease to be valid, 
As we increase $m$ from $m=-1/2+\delta$, the interactions increase and we can expect the quasi-particle description to cease to be valid beyond $m=0$. For that value $m=0$, and beyond $m>0$, the naive scaling dimension $\Delta_{\cO}$ of the fermion-{bilinear} $\cO_{\Delta_{\cO}}=\Psi\Psi$ is marginal or irrelevant and it is hard to see how the ultra-violet conformal theory can flow to a Fermi-liquid state, assuming that all vacuum state changes are caused by the condensation of bosonic operators. This intuition will be borne out by our results: when $m\geq 0$ the standard Fermi-liquid disappears.
A similar approach to describing fermionic quantum criticality \cite{Liu:2009dm} discusses the special case $m=0$ or $\Delta_{\Psi}=d/2$ in detail; other descriptions of the $m=0$ system are \cite{Lee:2008xf,Rey:2008zz}.

\paragraph*{The emergent Fermi liquid}
With an eye towards experiment we shall consider the AdS${}_4$ dual to a relativistic CFT${}_3$ in $d=2+1$ dimensions; see supporting material \cite{supp}. As we argue there, we do not know the detailed microscopic CFT nor whether a dual AdS with fermions as the sole U(1) charged field exists as a fully quantum consistent theory for all values of $m=\Delta_{\Psi}-d/2$, but the behavior of fermion spectral functions at a strongly coupled quantum critical point can be deduced nonetheless. Aside from $\Delta_{\Psi}$, the spectral function will depend on the dimensionless ratio $\mu_0/T$ as well as the $U(1)$ charge $g$ of the fermion; we shall set $g=1$ from here on, as we expect that only large changes away from $g=1$ will change our results qualitatively. We therefore study the system as a function of $\mu_0/T$ and $\Delta_{\Psi}$. We have drawn our approach in Fig. 1B: first we shall study the spectral behavior as a function of $\mu_0/T$ for fixed $\Delta_{\Psi}<3/2$; then we study the spectral behaviour as we vary the scaling dimension $\Delta_{\Psi}$ from $1$ to $3/2$ for fixed $\mu_0/T$ coding for an increasingly interacting fermion. Note that our set-up and numerical calculation necessitate a finite value of $\mu_0/T$: all our results are at non-zero $T$.  

\begin{figure}[t]
(A)\hspace*{3in}(B)\hspace*{2in}{}\\
\centering
\raisebox{.2in}{
\includegraphics[width=3.1in]{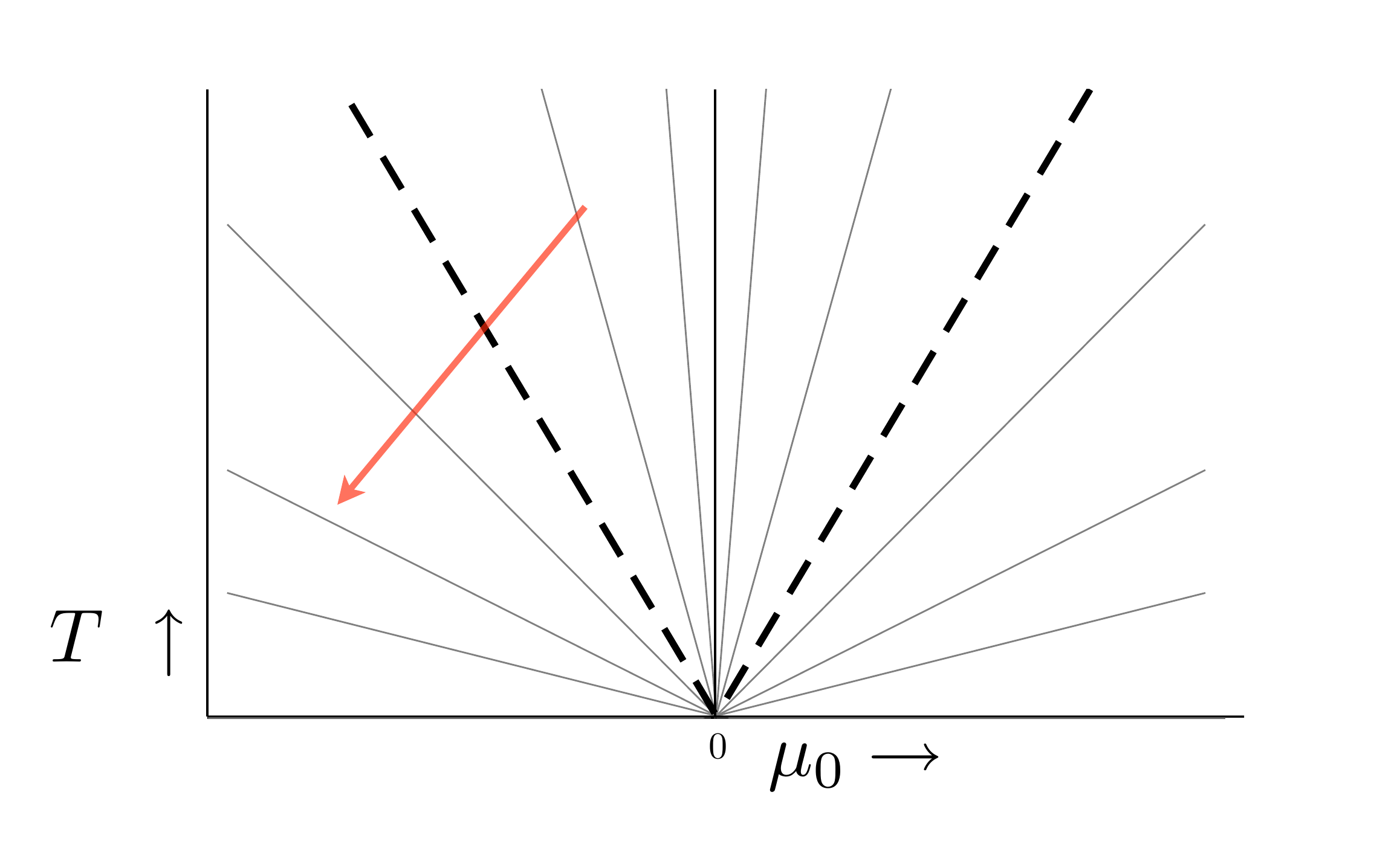}}
\includegraphics[width=2.7in]{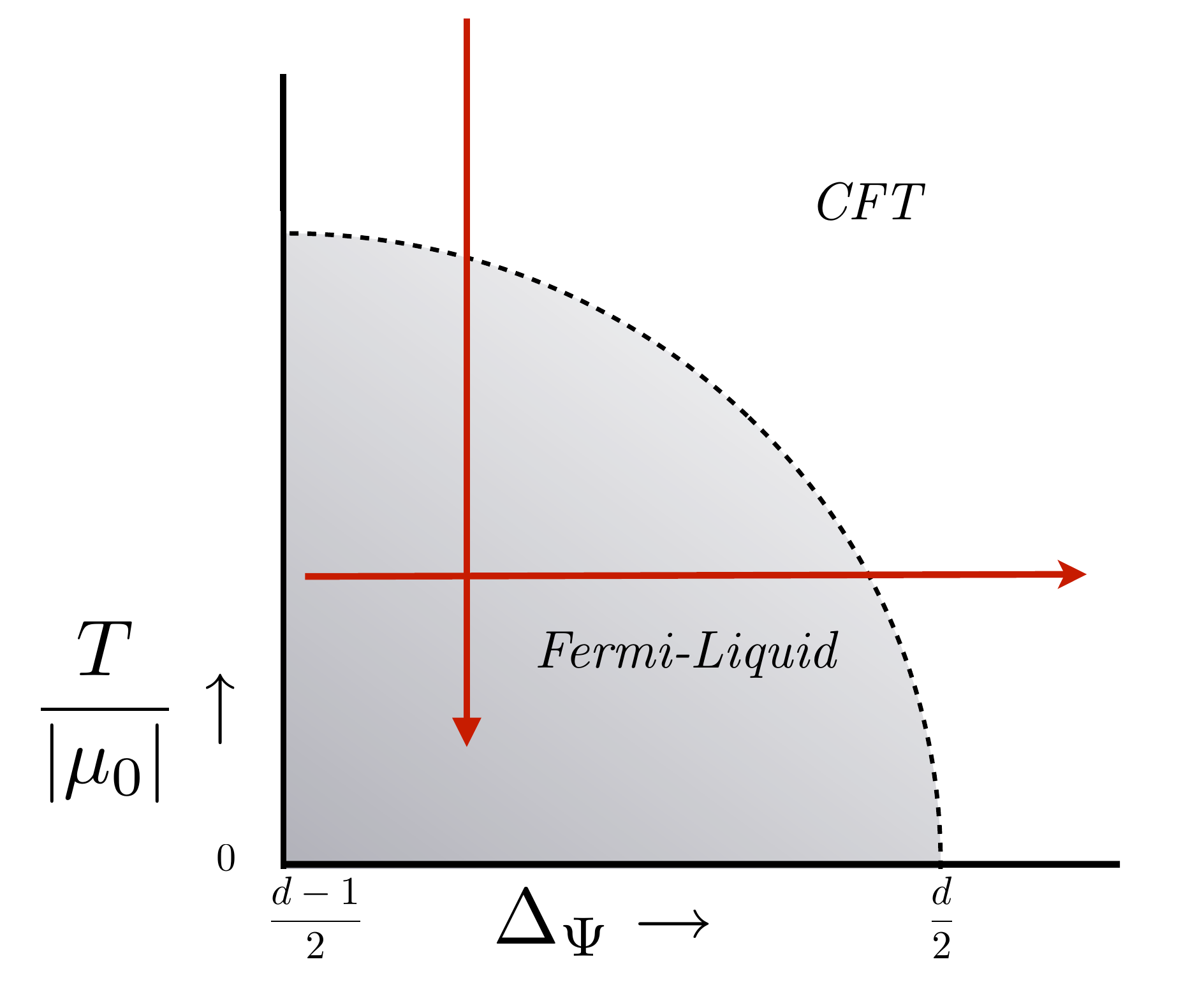}
\label{fig0} \caption{\em \small The phase diagram near a quantum-critical point. Gray lines depict lines of constant $\mu_0/T$: the spectral function of fermions is unchanged along each line if the momenta are appropriately rescaled. As we increase $\mu_0/T$ we crossover from the quantum-critical regime to the Fermi-liquid.
{\rm (B)} The trajectories in parameter space $(\mu_0/T, \Delta_{\Psi})$ studied here. We show the crossover from the quantum critical regime to the Fermi liquid by varying $\mu_0/T$ keeping $\Delta_{\Psi}$ fixed; we cross back to the critical regime varying $\Delta_{\Psi}\rar d/2$ for $\mu_0/T$ fixed. The boundary region is not an exact curve, but only a qualitative indication.}
\end{figure}
Our analysis starts near the reference point $\mu_0/T\rar 0$ where the long range behavior of the system is controlled by the quantum critical point (Fig. 1A). Here we expect to recover conformal invariance, as the system forgets about
any well-defined scales, and the spectral function should be controlled by the branchcut at $\ome=k$ in the Green's function (Eq.1)
%(\ref{eq:20})
: (a) For $\ome<k$ it should vanish, (b) At $\ome=k$ we expect a sharp peak which 
for $\ome \gg k $ scales as $\ome^{2\Delta_{\Psi}-d}$.  Fig. 2A shows this expected behavior of spectral function for three different
values of the momentum for a fermionic operator with weight $\Delta_{\Psi}=5/4$ computed from AdS${}_{4}$ following the set-up in the supporting material~\cite{supp}.

\begin{figure}[t]
%\begin{minipage}[t]{0.5\linewidth}
\centering
\parbox[t]{0.55\linewidth}{
(A)\\
(B)\includegraphics[width=3.5in]{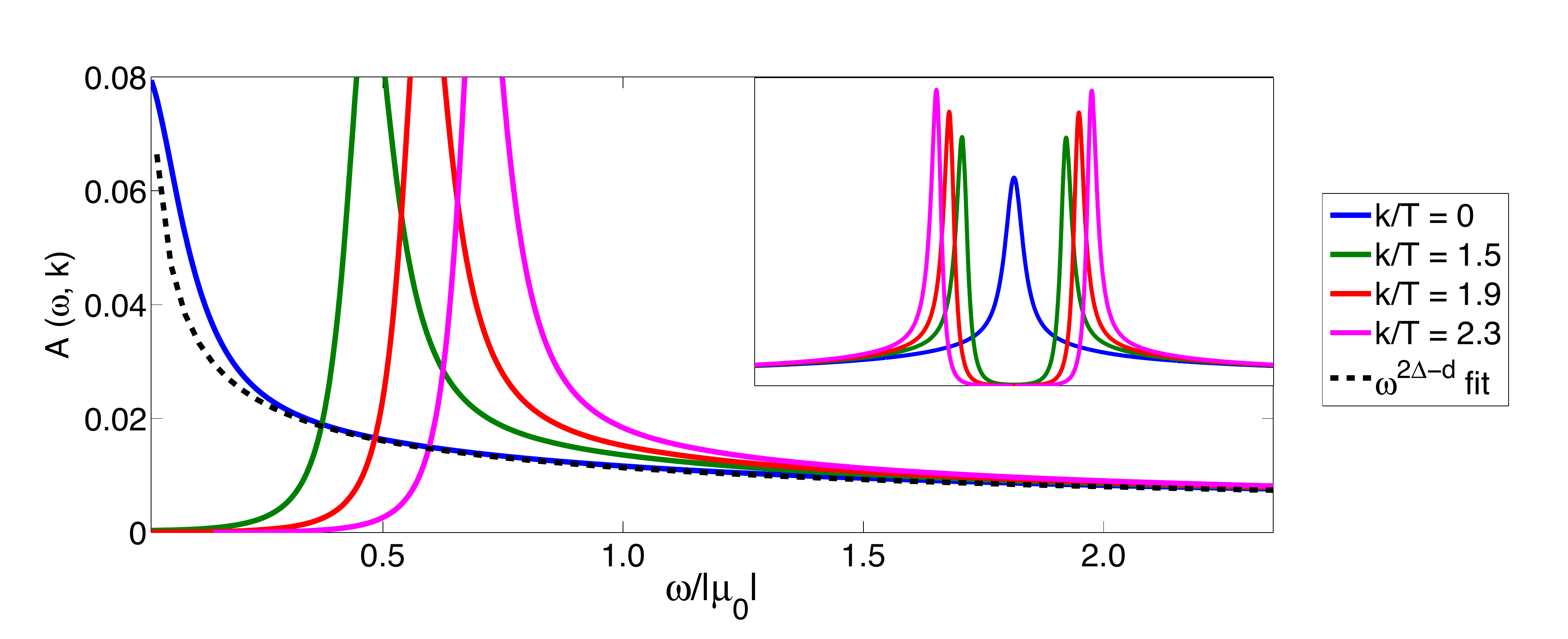}\\
%(C)
\hspace*{.28in}\includegraphics[width=3.6in]{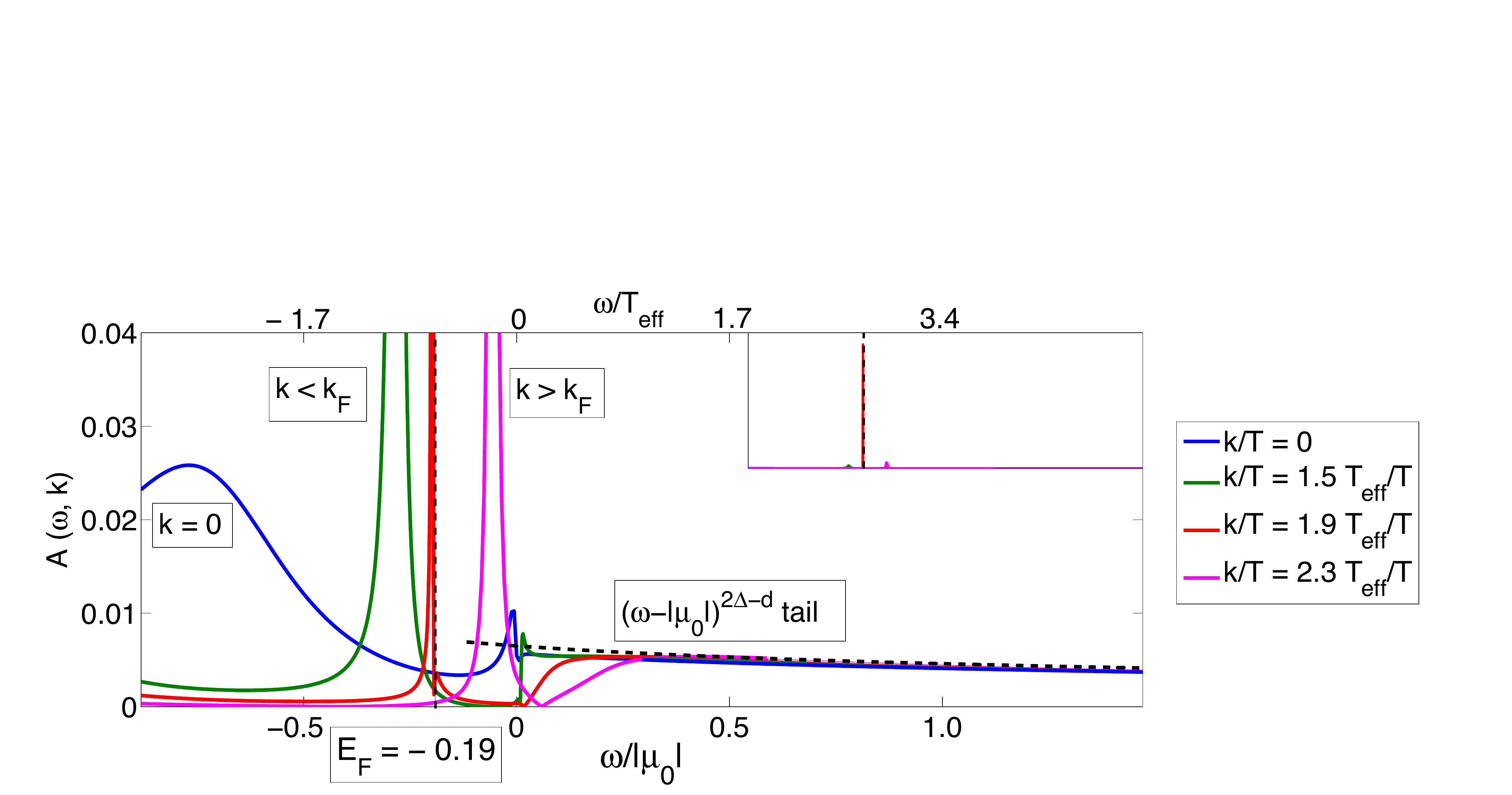}}
\parbox[t]{0.4\linewidth}{
(C)\\
~\\
\hspace*{.1in}\includegraphics[width=3in,height=2.25in]{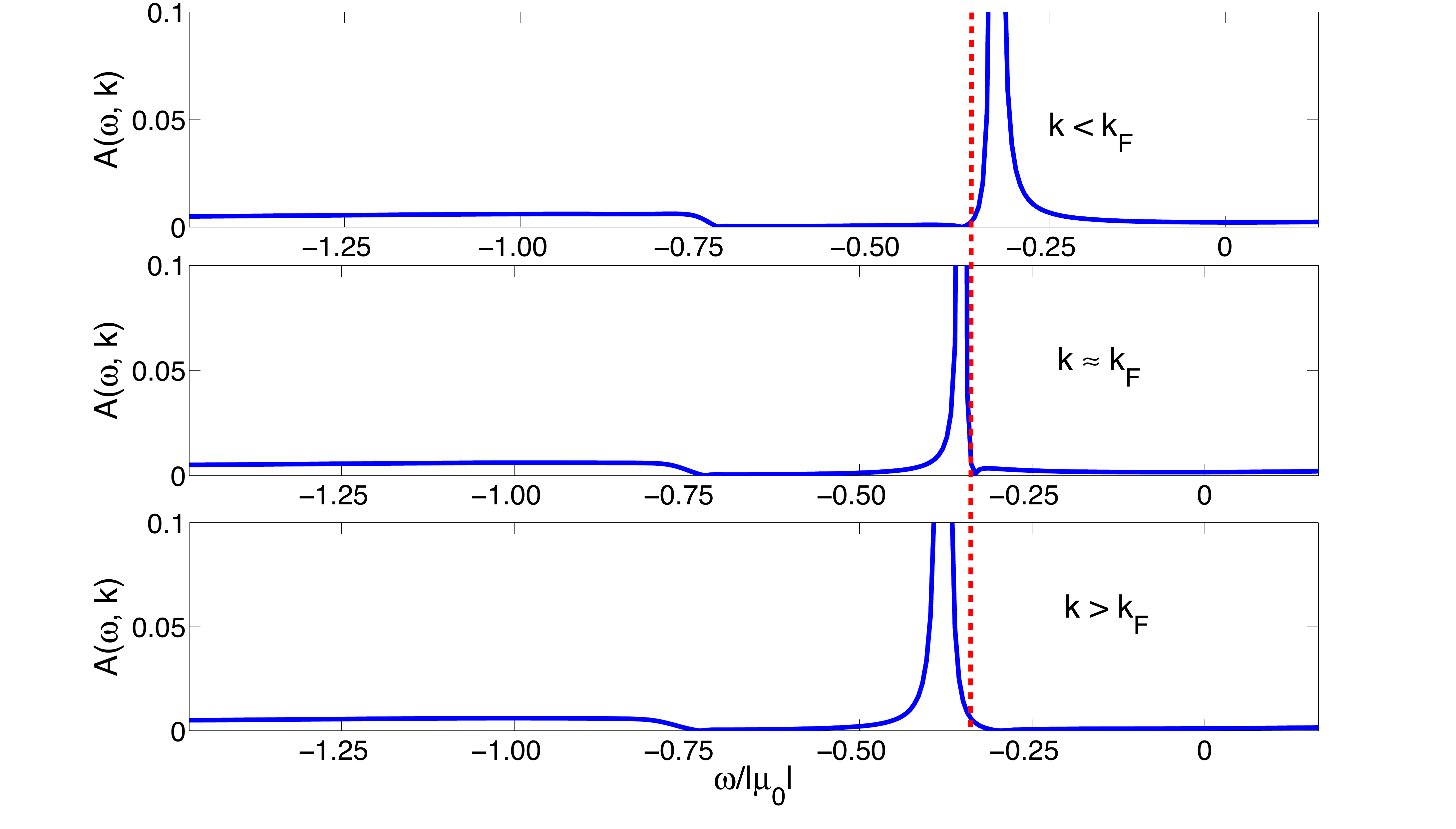}}
%\vspace*{-1in}

%}
%\label{fig1}
\caption{\em\small {\rm  (A)} The spectral function $A(\ome,k)$
for $\mu_0/T=0.01$ and $m=-1/4$. The spectral function has the asymptotic branch cut behavior of a conformal
field of dimension $\Delta_{\Psi}=d/2+m=5/4$:  it vanishes for $\ome<k$, save for a finite $T$ tail,
and for large $\ome$ scales as $\ome^{2\Delta_{\Psi}-d}$. 
{\rm (B)} The emergence of the quasiparticle peak as we change the chemical potential to $\mu_0/T=-30.9$ for the same 
value
$\Delta_{\Psi}=5/4$. The three displayed momenta $k/T$ are rescaled by a factor $T_{eff}/T$ for the most meaningful comparison with those in (A); see~\cite{supp}.  The insets show the full scales of the peak heights and the dominance of the quasiparticle peak for $k\sim k_F$.
{\rm (C)} Vanishing of the spectral function at $E_F$ for $\Delta_{\Psi}=1.05$ and $\mu_0/T =-30.9$. The deviation of the dip-location from $E_F$ is a finite temperature effect. It decreases with increasing $\mu_0/T$.}
\end{figure}
Turning on $\mu_0/T$ holding $\Delta_{\Psi}=5/4$ fixed, shifts the center location of the two branchcuts to an effective chemical potential $\ome=\mu_{eff}$; this bears out our expectation that the $U(1)$ chemical potential induces a finite fermion density. While the peak at the location of the negative branchcut $\ome\sim\mu_{eff}-k$ stays broad, the peak at the other branchcut $\ome\sim\mu_{eff}+k$ sharpens distinctively as the size of $\mu_0/T$ is increased (Fig. 2B). {We shall identify this peak with the quasiparticle of the Fermi liquid and its appearance as the crossover between the quantum-critical and the Fermi-liquid regime.} The spectral properties of the Fermi liquid are very well known and display a number of uniquely identifying characteristics \cite{Landau,Schulz-et-al}. If this identification is correct, all these characteristics must be present in our spectra as well. 
 \begin{enumerate}
 \item The quasiparticle peak should approach a delta function at the Fermi momentum $k=k_F$. In Fig. 2B we see the peak narrow as we increase $k$, peak, and broaden as we pass $k\sim k_F$ (recall that $T=0$ is outside our numerical control and the peak always has some broadening).  In addition the spectrum should vanish identically at the Fermi-energy $A(\ome=E_F,k)=0$, independent of $k$. This is shown in Fig. 2C.

\item The quasiparticle should have linear dispersion relation near the Fermi energy with a renormalized Fermi velocity $v_F$ different than the underlying relativistic speed $c=1$. In Fig. 3 we plot the maximum of the peak $\ome_{max}$ as a function of $k$. At high $k$ we recover the linear dispersion relation \mbox{$\ome =|k|$} underlying the Lorentz invariant branchcut in Eq.1.
%~(\ref{eq:20}) 
%\comment{Equation 1}.
 Near the Fermi energy/Fermi momentum however, this dispersion relation changes to a slope 
\mbox{$v_F \equiv \lim_{\ome\rar E_F,k\rar k_F}(\ome-E_F)/(k-k_F)
$} 
clearly less than one. Importantly, we learn that the Fermi Energy $E_F$ is not located at zero-frequency. Recall, however, that the AdS chemical potential $\mu_{0}$ is the bare $U(1)$ chemical potential in the CFT. This is confirmed in Fig. 3 from the high $k$ behavior: its Dirac point is $\mu_0$. On the other hand, the chemical potential felt by the IR fermionic degrees of freedom is renormalized to the value $\mu_F=\mu_0-E_F$. As is standard, the effective energy $\tilde{\ome}=\ome-E_F$ of the quasiparticle is measured with respect to $E_F$.  
\begin{figure}
\centering
\includegraphics[width=4.75in]{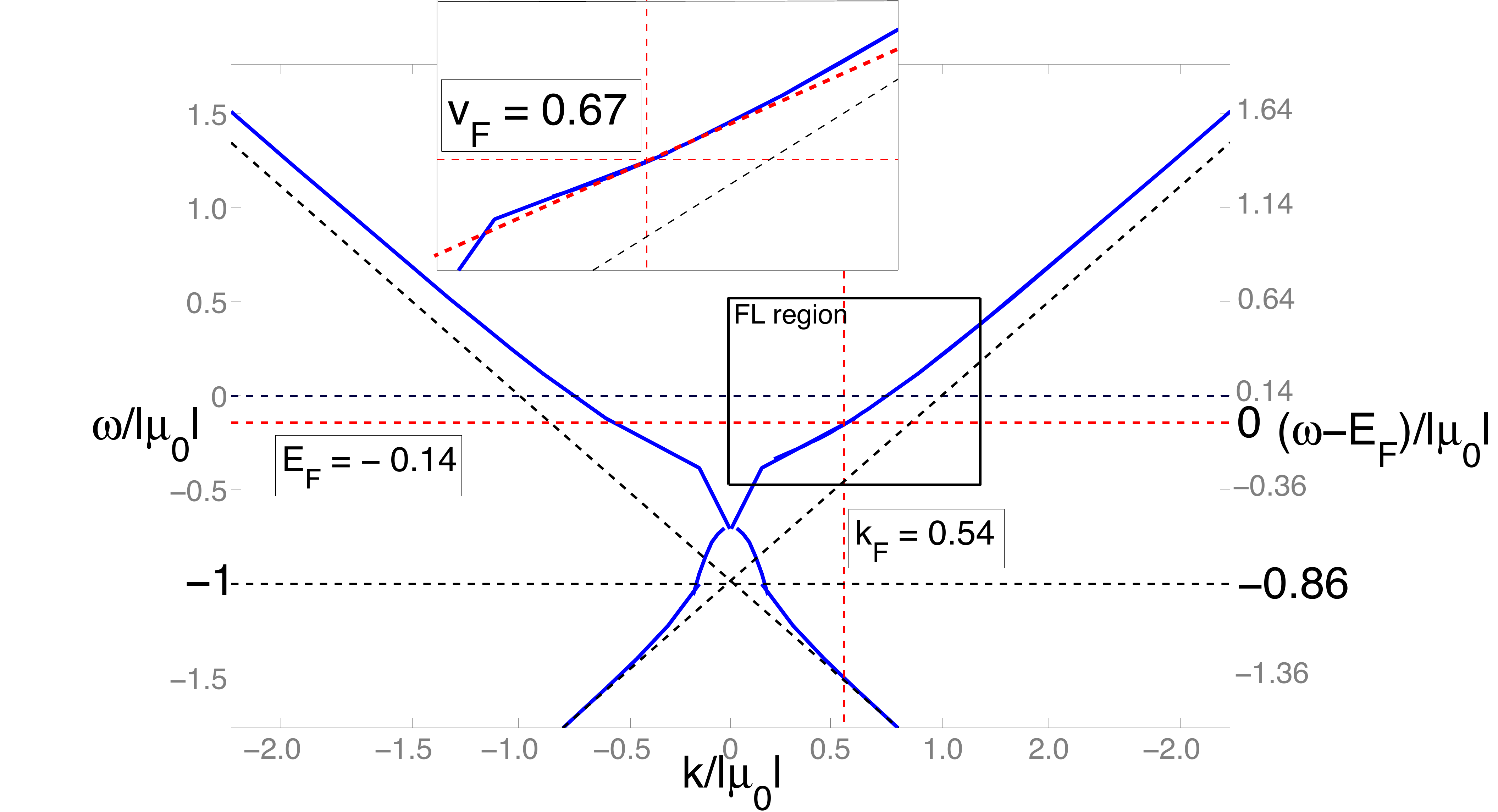}
\caption{\small\it Maxima in the spectral function as a function of $k/\mu_0$ for $\Delta_{\Psi}=1.35$ and $\mu_0/T=-30.9$. Asymptotically for large $k$ the negative $k$ branch cut recovers the Lorentz-invariant linear dispersion with unit
velocity, but with the zero shifted to $-\mu_{0}$. The peak location of the positive $k$ branch cut that changes into the quasiparticle peak changes significantly. It gives the dispersion relation of the quasiparticle near $(E_F, k_F)$. The change of the slope from unity shows renormalization of the Fermi velocity. This is highlighted in the inset. 
Note that the Fermi energy $E_F$ is not located at $\ome_{AdS}=0$. 
The AdS calculation visualizes the renormalization of the bare UV chemical potential $\mu_0=\mu_{AdS}$ to the effective chemical potential $\mu_{F}=\mu_0-E_F$ felt by the low-frequency fermions.}
\end{figure}

\item  At low temperatures Fermi-liquid theory predicts the
width of the quasiparticle peak to grow quadratically with temperature. Fig. 4A, 4B show this distinctive behavior up to a critical temperature 
%$(T/T_{eff})^2 \sim 0.16$ or equivalenty $(\mu_0/T)_c \sim 3.6$.
$T_c/\mu_0\sim 0.16$.
This temperature behavior directly follows from the fact that imaginary part of the self-energy $\Sig(\ome,k)=\ome-k-({\rm Tr} i\gam^0 G(\ome,k))^{-1}$ should have no linear term when expanded around $E_F$: Im$\Sig(\ome,k)\sim (\ome-E_F)^2 +...$. This is shown in Fig. 4C, 4D.

\end{enumerate}
These results give us confidence that we have identified the characteristic quasiparticles at the Fermi surface of the Fermi liquid emerging from the quantum critical point.

%\paragraph*{Fermi-liquid evolves}

Let us now discuss how this Fermi-liquid evolves when we increase the  bare $\mu_0$ (Fig. 5). 
\begin{figure}
  \centering
\parbox[t]{2.7in}{
%\begin{center}
%\vspace{-2.65in}
(A)\\
(B)\includegraphics[width=2.85in]{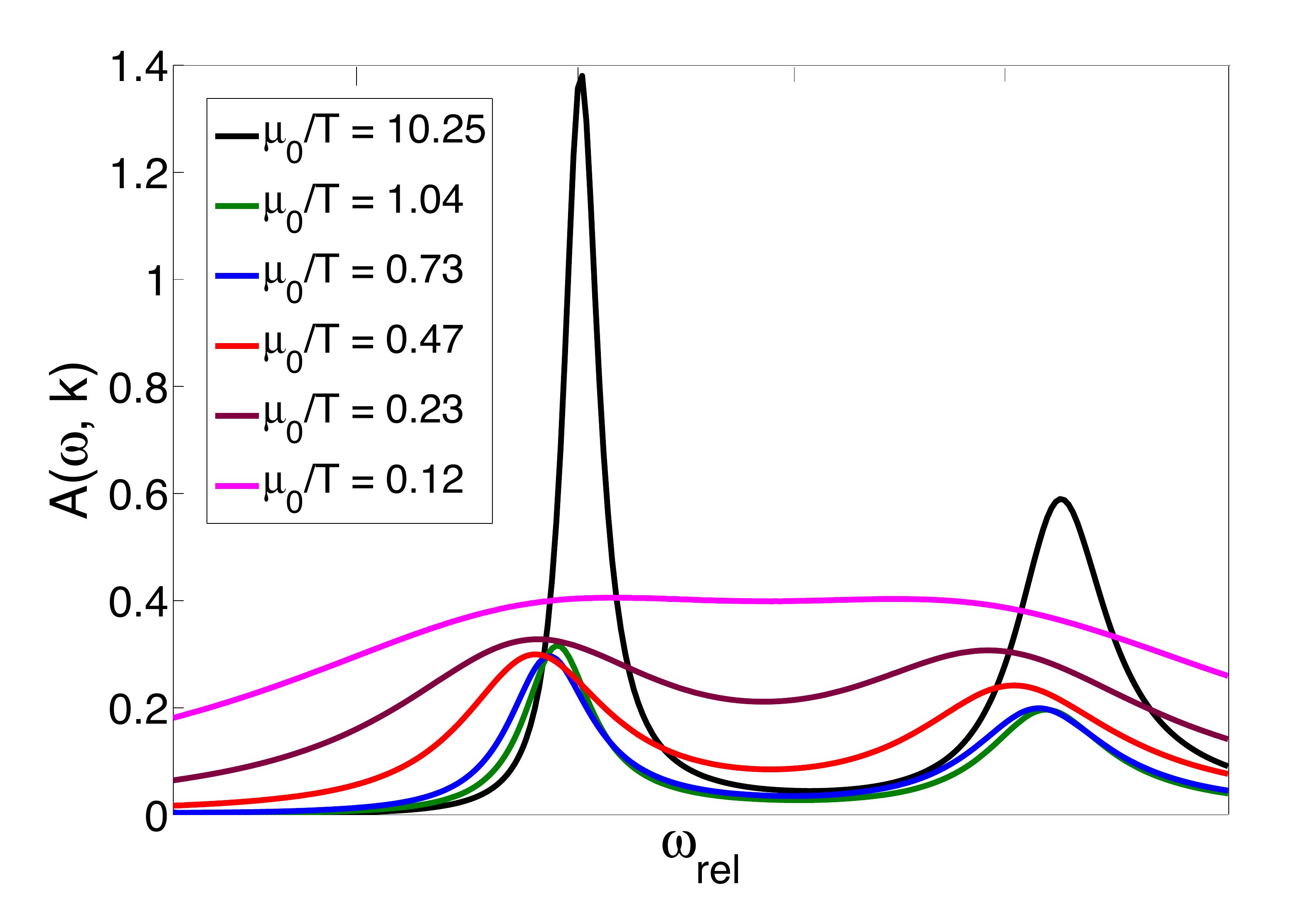}\\[.1in]
\hspace*{.42in}\includegraphics[height=1.94in]{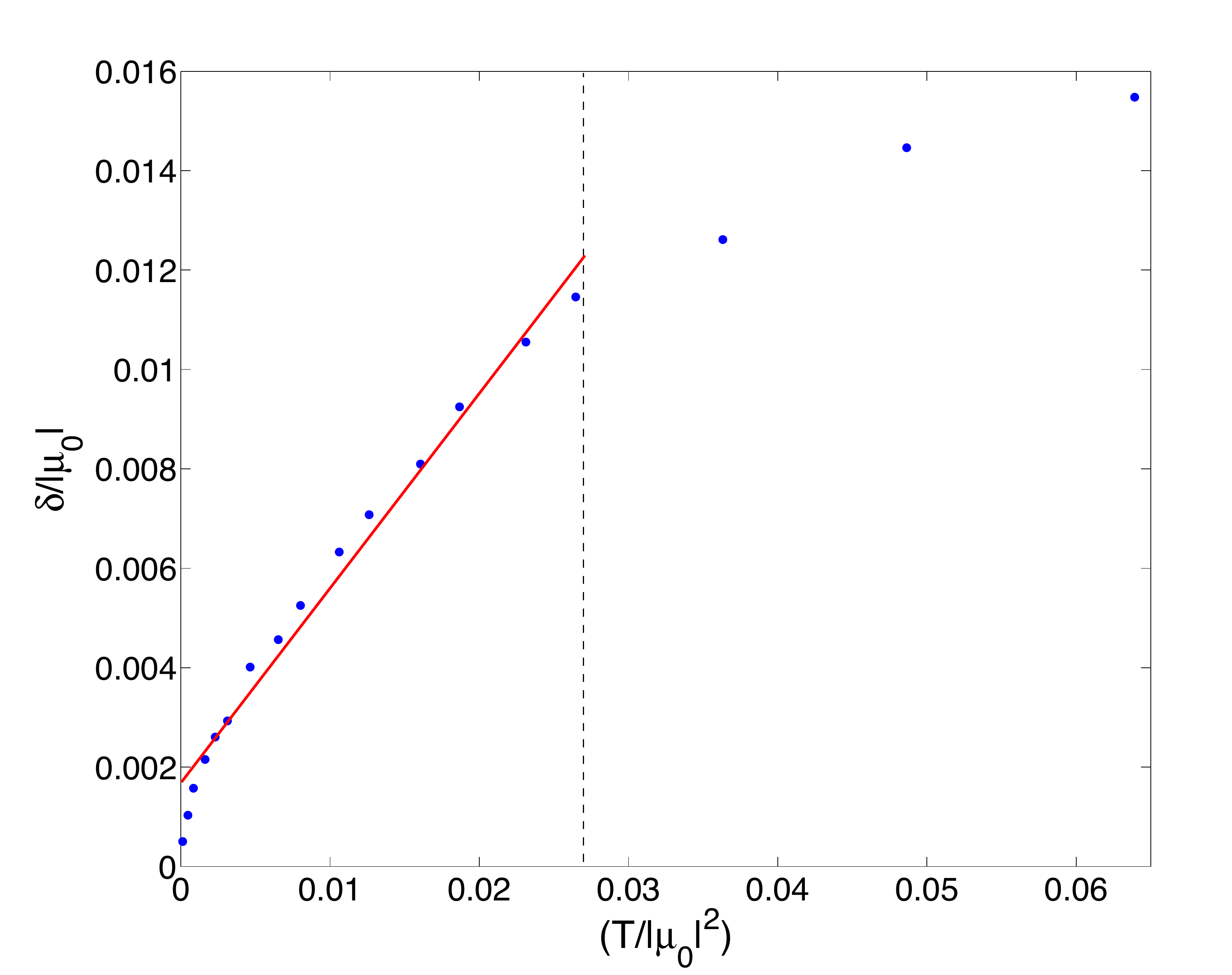}
%\end{center}
}~~~~~~
\parbox[t]{3.15in}{
(C)\\[-.05in]
(D)\includegraphics[width=2.6in]{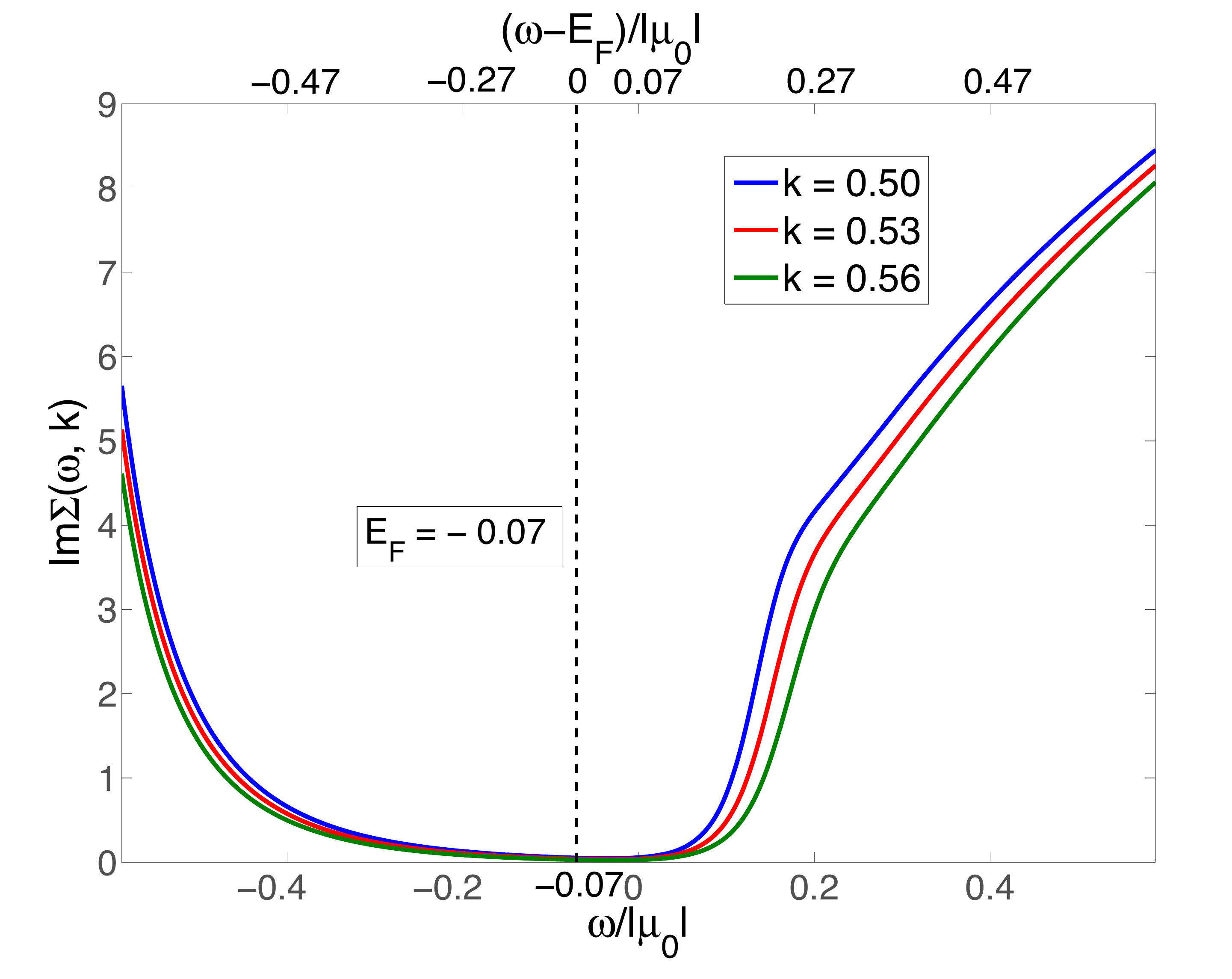}\\
\hspace*{.12in}\includegraphics[width=2.75in]{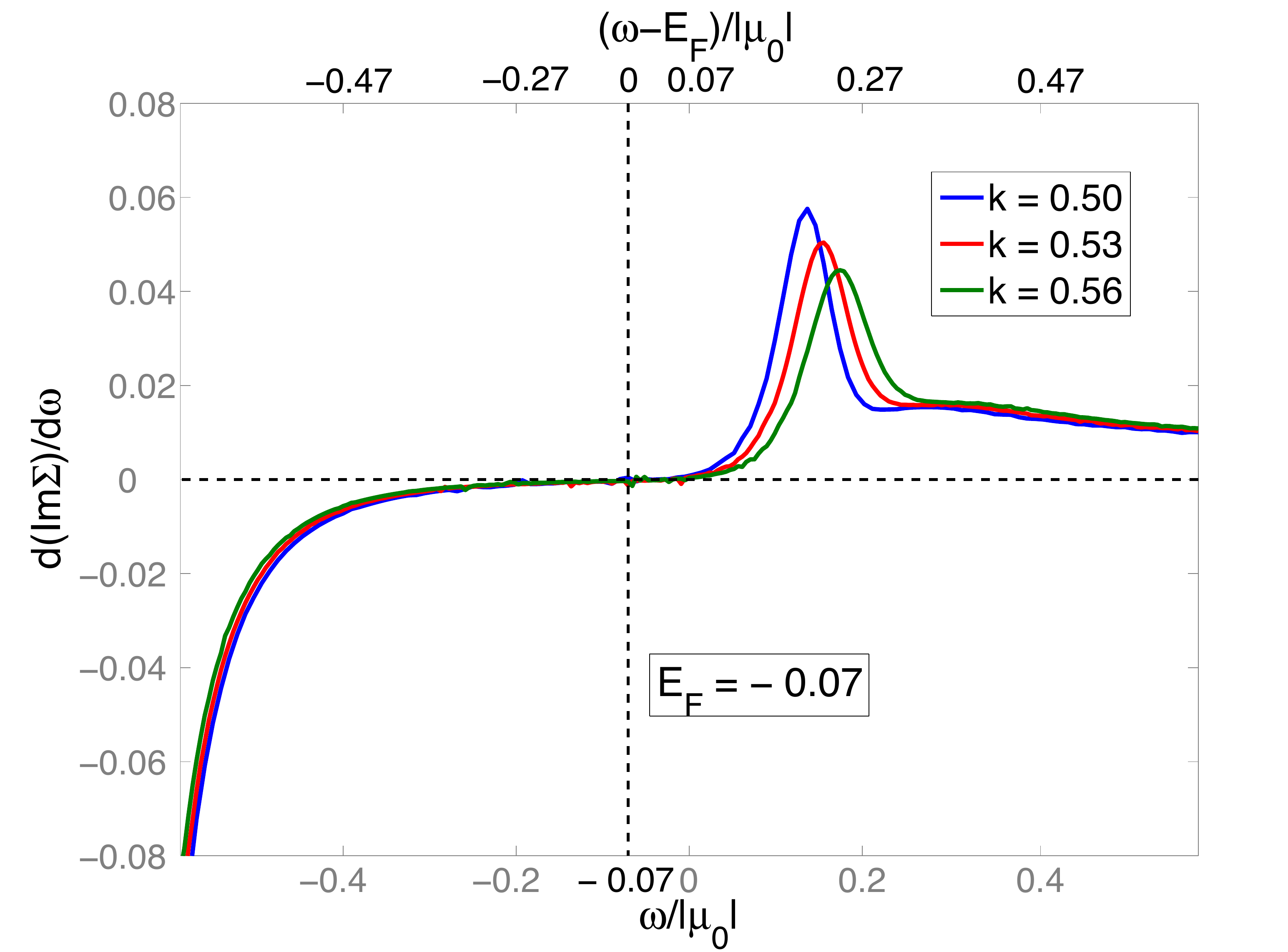}}
\caption{\small \it
{\rm (A)} Temperature dependence of the quasiparticle peak for $\Delta_{\Psi}=5/4$ and $k/k_F \simeq 0.5$; all curves have been shifted to a common peak center. {\rm (B)} The quasiparticle peak width $\delta \sim {\rm Re}\Sig(\ome,k=k_F)$ % in $\ome/T_{eff}$ at $k=k_F$
  for $\Delta_{\Psi}=5/4$ as a function of $T^2$: it reflects the expected behavior $\delta \sim T^2$ up to a critical temperature $T_c/\mu_0$, beyond which the notion of a quasiparticle becomes untenable. {\rm (C)} The imaginary part of the self-energy $\Sig(\ome,k)$ near $E_F,~k_F$ for $\Delta_{\Psi}=1.4,~\mu_0/T=-30.9$ . The defining ${\rm Im} \Sig(\ome,k)\sim (\ome-E_F)^2+\ldots$-dependence for Fermi-liquid quasiparticles is faint in  panel {\rm (C)} but obvious in panel {\rm (D)}. It shows that the intercept of $\pa_{\ome}{\rm Im} \Sig(\ome,k)$ vanishes at $E_F,k_F$.
}
\end{figure}
Similar to the fermion chemical potential $\mu_F$, the fundamental control parameter of the Fermi-liquid, the fermion density $\rho_F$, is not directly related to the AdS $\mu_0$. We can, however, infer it from the Fermi-momentum $k_F$  that is set by the quasiparticle pole via Luttinger's theorem $\rho_F \sim k_F^{d-1}$. 
The more illustrative figure is therefore Fig. 5B which shows the quasiparticle characteristics as a function of $k_F/T$. We find that the quasiparticle velocities decrease slightly with increasing $k_F$, rapidly %asymptoting to 
levelling off to a finite constant less than the relativistic speed. Thus the quasiparticles become increasingly heavy as their mass $m_F \equiv  k_F/v_F$ asymptotes to linear growth with $k_F$. The Fermi energy $E_F$ also shows linear growth. Suppose the heavy Fermi-quasiparticle system has the underlying canonical non-relativistic dispersion relation $E=k^2/(2m_F)=k_F^2/(2m_F)+ v_F(k-k_F)+...$, then the observed Fermi energy $E_F$ should equal the renormalized Fermi-energy $E^{(ren)}_F \equiv k^2_F/(2m_F)$. Fig. 5B shows that these energies $E_F$ and $E_F^{(ren)}$ track each other remarkably well. We therefore infer that the true zero of energy of the Fermi-quasiparticle is set by the renormalized Fermi-energy as deduced from the Fermi-velocity and -momentum.  
 
\begin{figure}
(A)\hspace*{3.5in}(B)\hspace*{2in}{}\\
  \centering
%\parbox[t]{2.6in}{
\includegraphics[width=3in]{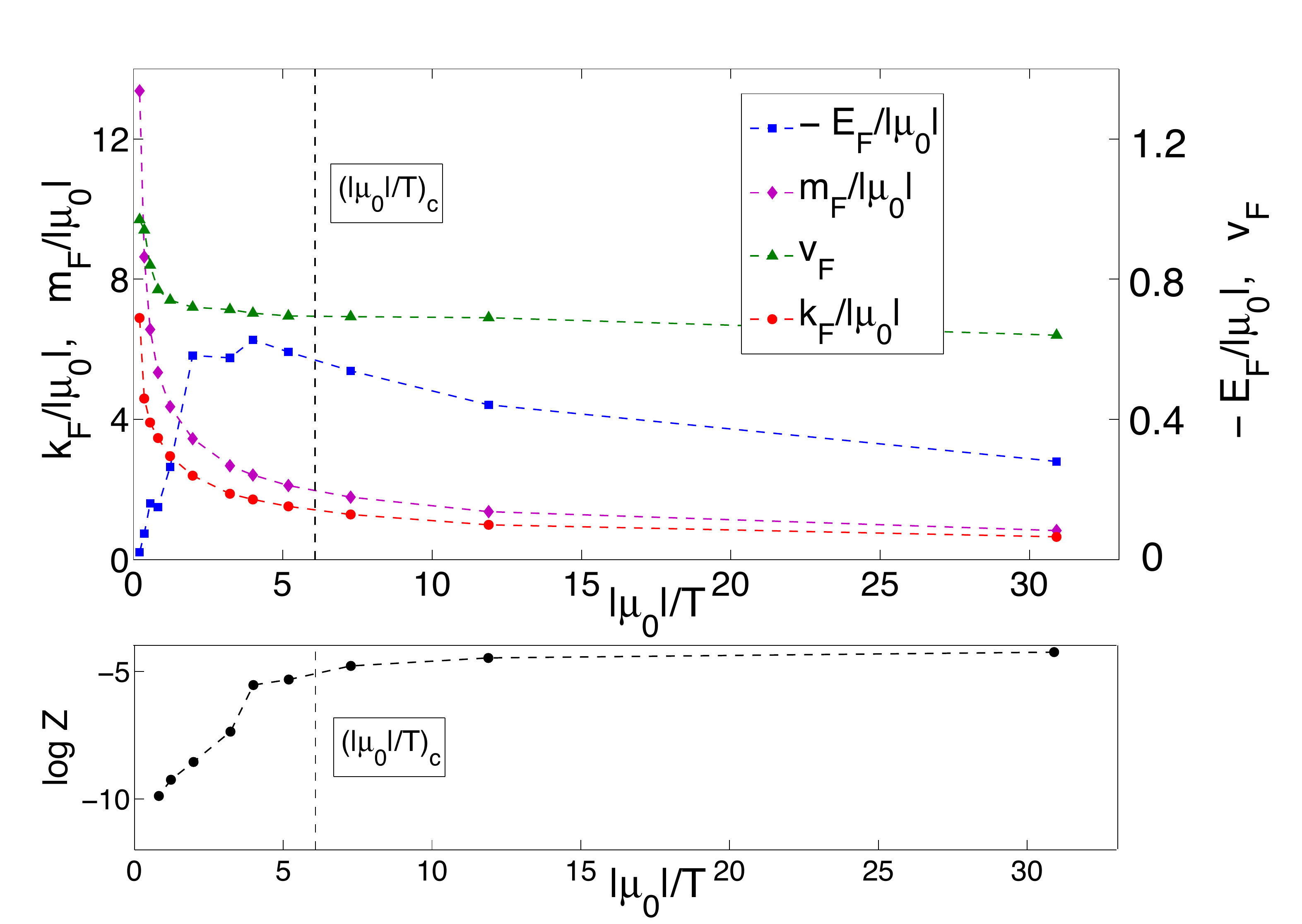}
\includegraphics[width=2.9in]{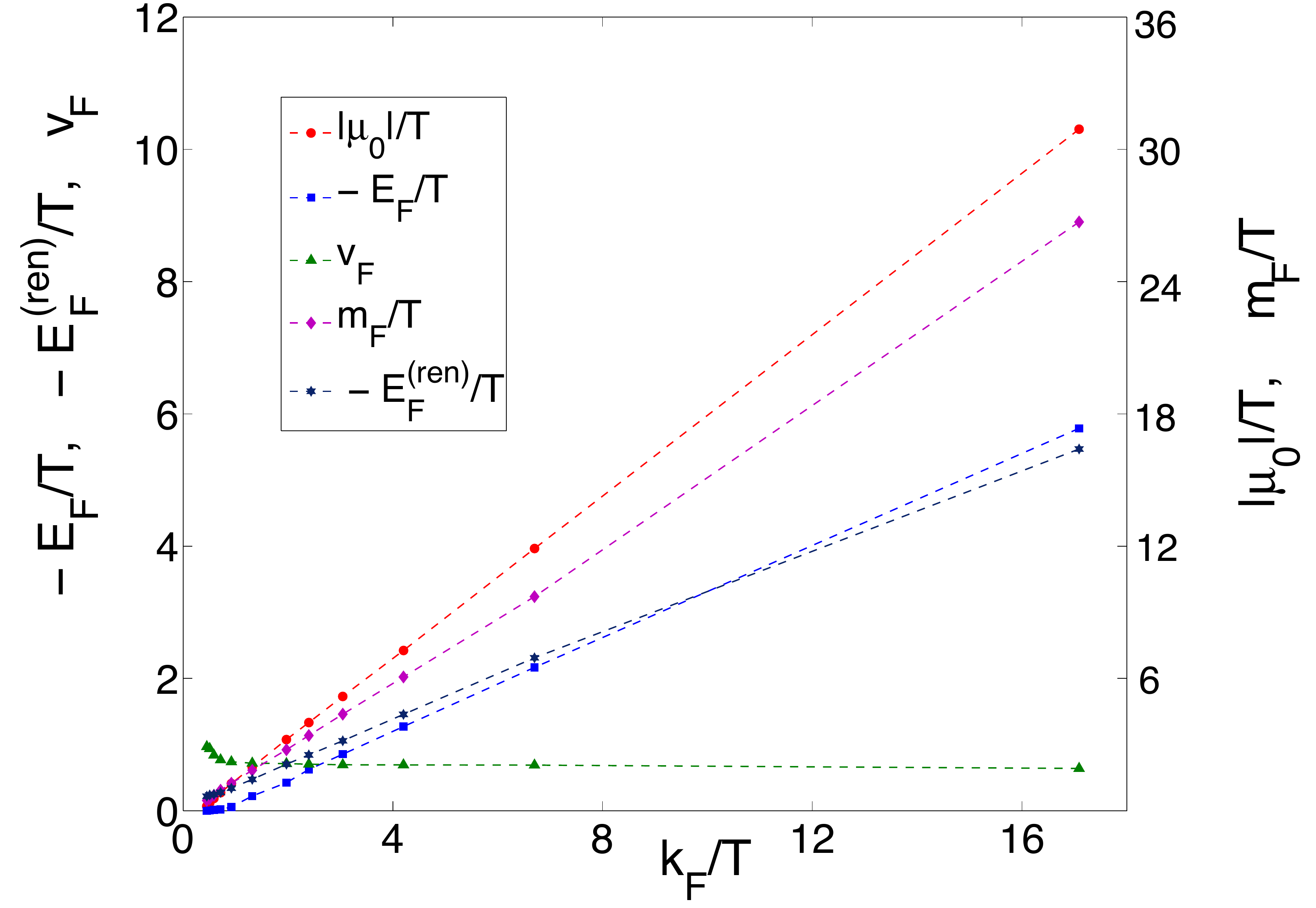}
%}
%\includegraphics[width=3in]{temp}
%\includegraphics[width=3in]{Fig4}
\caption{\small \it
The quasiparticle characteristics as a function of $\mu_0/T$ for $\Delta_{\Psi}=5/4$. Panel {\rm (A)} shows the change of $k_F, v_F, m_F, E_F$ and the pole strength $Z$ (the total weight between half-maxima) as we change $\mu_0/T$. Beyond a critical value $(\mu_0/T)_c$ we lose the characteristic $T^2$ broadening of the peak and there is no longer a real quasiparticle, though the peak is still present. For the Fermi-liquid $k_F/T$ rather than $\mu_0/T$ is the defining parameter. We can invert this relation and panel {\rm (B)} shows the quasiparticle characteristics as a function of $k_F/T$. Note the linear relationships of $m_F,E_F$ to $k_F$ and that the renormalized Fermi energy $E^{(ren)}\equiv k_F^2/(2m_F)$ matches the empirical value $E_F$ remarkably well.
}
\end{figure}
Although the true quasiparticle behavior disappears at $T>T_c$, Fig. 5A indicates that in the limit $k_F/T \rightarrow 0$ the quasiparticle pole strength vanishes, $Z_k \rightarrow 0$, while the Fermi-velocity $v_F$ remains finite; $v_F$ approaches the bare velocity $v_F=1$.  This is seemingly at odds with the ‘heavy’ Fermi liquid wisdom $Z_k \sim m_{micro}/m_F=m_{micro}v_F/k_F$. The resolution is the restoration of Lorentz invariance at zero density. From general Fermi liquid considerations it follows that $v_F  = Z_k ( 1 + \left. \partial_k {\rm Re}\Sigma\right|_{E_F,k_F})$ and $Z_k = 1 / ( 1- \left.\partial_{\omega} {\rm Re} \Sigma\right|_{E_F,k_F})$ where $\partial_{k,\omega} {\rm Re}\Sigma$ refers to the momentum and energy derivatives of the real part of the fermion self-energy $\Sig(\ome,k)$ at $k_F,E_F$. Lorentz invariance imposes $\partial_{\omega} \Sigma' =- \partial_k \Sigma'$ which allows for vanishing $Z_k$ with $v_F\rar 1$. Interestingly, the case has been made that such a relativistic fermionic behavior might be underlying the physics of cuprate high $T_c$ superconductors \cite{randeria}.  

Finally, we address the important question what happens
when we vary the conformal dimension $\Delta_{\Psi}$ of the fermionic operator. 
\begin{figure}
\centering
\includegraphics[width=4.75in]{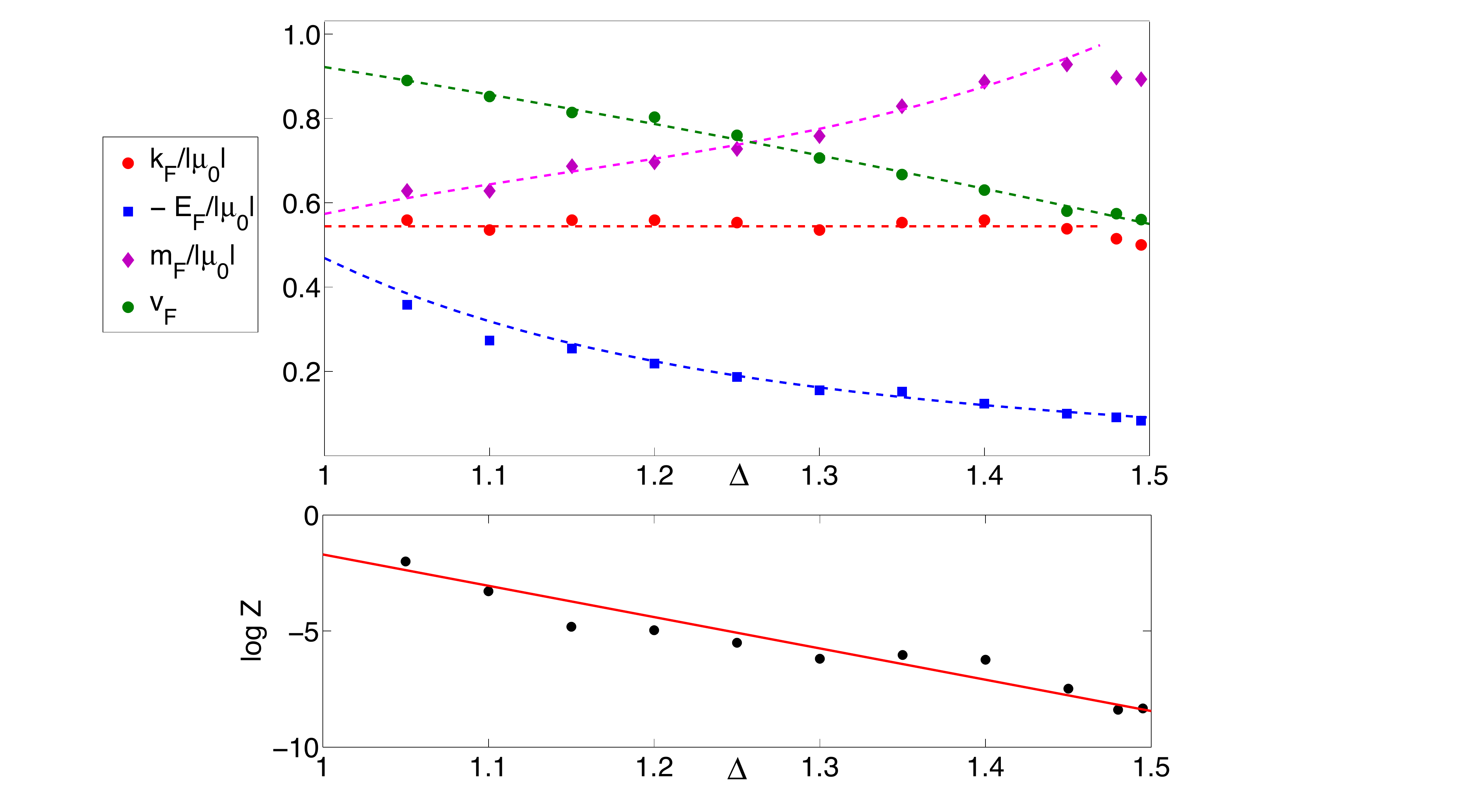}
\caption{\small \it 
The quasiparticle characteristics as a function of the
Dirac fermion mass $-1/2<m<0$ corresponding to $1 <\Delta_{\Psi} <3/2$ for $\mu_0/T=-30.9$.
The upper panel shows the independence of $k_F$ of the mass. This indicates Luttinger's theorem if the anomalous dimension $\Delta_{\Psi}$ is taken as an indicator of the interaction strength. Note that $v_F, E_F$ both asymptote to finite values as $\Delta_{\Psi} \rar 3/2$. The lower panel shows the exponential vanishing pole strength $Z$ (the integral between the half-maxima) as $m\rar 0$.
}
\end{figure} 
Fig. 6 shows that the Fermi momentum $k_F$ stays constant as we increase $\Delta_{\Psi}$.  This completes our identification of the new phase as the Fermi-liquid: it indicates that the AdS dual obeys Luttinger's theorem, {if} we can interpret the conformal dimension of the fermionic operator as a proxy for the interaction strength. We find furthermore that the quasiparticle pole strength vanishes as we approach $\Delta_{\Psi}=3/2$. This confirms our assumption made earlier that it is essential to study the system for $\Delta_{\Psi}<d/2$ and that the point $\Delta_{\Psi}=d/2$ where the naive fermion bilinear becomes marginal signals the onset of a new regime. Because the fermion bilinear is marginal at that point this ought to be an interesting regime in its own right and we refer to the recent article \cite{Liu:2009dm} for a discussion thereof~\cite{footnote2}.
% \footnote{Note that the $m=0$ spectral peak discussed in the article \cite{Liu:2009dm} is therefore not the peak we identified with the quasiparticle state. See supporting material.} 
Highly remarkable is that the pole strength vanishes in an {exponential} fashion rather than the anticipated algebraic behavior \cite{Senthil,Krueger}. This could indicate that an essential singularity governs the critical point at $\Delta_{\Psi}=d/2$ and we note that such a type of behavior was identified by Lawler {et al.} in their analysis of the Pomeranchuk instability in $d=2+1$ dimensions using the ‘Haldane patching’ bosonization procedure \cite{essential}.  Interestingly this finite $\mu_0/T$ transition as we vary $\Delta_{\Psi}$ has no clear symmetry change, similar to \cite{Krueger}. However, this may be an artifact of the fact that our theory is not quantum mechanically complete \cite{supp}. Note also that  the quasiparticle velocity and the renormalized Fermi energy $E_F=v_F(k-k_F)-E$ stay finite at the $\Delta_{\Psi}=3/2$ transition with $Z\rar 0$, which could indicate an emergent Lorentz invariance for the reasons discussed in the previous paragraph.

%\subsection{Novel unexpected features of the AdS Fermi liquid}
%section{Discussion and Outlook}

\paragraph*{Concluding remarks}
%In conclusion, 
We have presented evidence that the AdS dual description of strongly coupled field theories can describe the emergence of the Fermi-liquid from a quantum critical state --- both as a function of density and interaction strength as encoded in the conformal dimension of the fermionic operators.  From the AdS gravity perspective, it was unclear whether this would happen. Sharp peaks in the CFT spectral function correspond to so-called quasinormal modes of black holes \cite{Kovtun:2005ev}, but Dirac quasinormal modes have received little study (see e.g. \cite{Cho:2007de}). It is remarkable that the AdS calculation processes the Fermi-Dirac statistics essential to the Fermi-liquid correctly. This is manifested by the emergent renormalized Fermi-energy and the validity of Luttinger's theorem. The AdS gravity computation, however, is completely classical without explicit quantum statistics, although we do probe the system with a fermion.
It would therefore be interesting to fully understand the AdS description of what is happening, in particular how the emergent scales $E_F$ and $k_F$ feature in the geometry. An early indication of such scales was seen in \cite{Shieh:2008nf,Rozali:2007rx} in a variant of the story that geometry is not universal in string theory: the geometry depends on the probe used and different probes experience different geometric backgrounds. The absence of these scales in the general relativistic description of the AdS black hole could thus be an artifact of the Riemannian metric description of spacetime. 

Regardless of these questions, AdS/CFT has shown itself to be an powerful tool to describe finite density Fermi systems. The description of the emergent Fermi liquid presented here 
argues that AdS/CFT is uniquely suited as a computational device for field-theory problems suffering from fermion sign-problems. AdS/CFT represents a rich mathematical environment and
%and the results presented here provide 
a new approach to investigate qualitatively and quantitatively important questions in quantum many-body theory at finite fermion density.
%: fermionic superconductivity and transport, BCS pairing, Quantum Hall physics, emergent Lorentz invariance, etc. as the recent articles \cite{Ammon:2009fe,Erdmenger:2008yj,Kim:2008bv,Brodie:2000yz,Davis:2008nv,Gubser:2008wz, Kachru:2008yh}  and many mentioned here have already started to explore.
% and in ways that are still very mysterious manages to overcome the fermion-sign problem that has plagued theoretical condensed matter physics for so long. 

%\medskip
\appendix
\setcounter{section}{18}
\renewcommand{\theequation}{\thesection.\arabic{equation}}

\section{Supporting material}
\subsection{The AdS set up and AdS/CFT Fermion Green's functions.}
\setcounter{equation}{0}
\label{sec:ads-set-up}

The deviation from the strongly coupled 2+1 dimensional quantum critical point from which we wish to see the Fermi surface emerge is characterized by a temperature and
background $U(1)$ chemical potential. The phenomenological AdS dual to such a finite-temperature system with chemical potential is a charged AdS${}_{4}$ black hole. 
Including fermionic excitations, this system is described by 
the minimal action
\begin{eqnarray}
  \label{eq:3}
  S_{bulk} = \frac{1}{2\kap_4^2} \int d^4x
  \sqrt{-g}\left[R+\frac{6}{L^2}+L^2\left(-\frac{1}{4}F^2 -
      \bar{\Psi}e_A^M\Gamma^AD_M\Psi -m\bar{\Psi}\Psi\right)\right].
\end{eqnarray}
Here $e_A^M$ is the inverse vielbein,
$\Gamma^A=\left\{\gam^{a},\gam^4\right\}$ are 4d Dirac
matrices obeying $\{\Gamma^A,\Gamma^B\}= 2\eta^{AB}$ (hermitian except
$\Gamma^0$), and $\Psi$ is a four-component Dirac spinor with
$\bar{\Psi}= \Psi^{\dag}i\Gam^0$.
This spinor is charged under a $U(1)$ gauge field and the covariant derivative equals
\begin{equation}
D_M\Psi =\left(\pa_M+\frac{1}{8}\ome_M^{AB}[\Gamma_A,\Gamma_B]+igA_M\right)\Psi~.
\end{equation}
On its own this action is not a consistent quantum theory. It must be embedded
in a string dual, e.g. for appropriate choices of $m$ and $g$ it is a subsector of the $\cN=8$ AdS${}_4 \times S_7$ dual to
the conformal fixed point of 
large $N_c$, $d=3$ $\cN=8$ SYM and generically such a completion will have a number of $U(1)$ charged fields in addition to the fermions. For our considerations, specifically the
two point function of fermions,
the quantum completion is not relevant. At leading order in the gravitational coupling constant,
the action (\ref{eq:3}) will yield the same two-point correlators independently of the non-linear supergravity couplings. It does mean, that we cannot equate the $U(1)$ chemical potential $\mu_0$ directly with the density of fermions $\mu_F$ as we emphasize in the main article. 

The charged AdS${}_4$ black hole is a solution to the equations of motion of this action.  In a gauge where $A_z=0$
and $A_0$ is regular at the horizon the metric and gauge potential are given by \cite{Romans:1991nq,Hartnoll:2007ai}
\begin{eqnarray}
  \label{eq:1}
  ds^2 &=& \frac{L^2\alp^2}{z^2}\left(-f(z)dt^2+dx^2+dy^2\right)+\frac{L^2}{z^2}\frac{dz^2}{f(z)} ~,\non
  A_{0} &=& 2q\alp(z-1)~, \nonumber \\[0.1in]
  f(z) &=& (1-z)(z^2+z+1-q^2z^3)~.  
\end{eqnarray}
For $z\rar 0$ the metric asymptotes to AdS${}_4$ in Poincar\'{e} coordinates with the boundary at $z=0$ and there is a black hole horizon at the first zero, $z=1$, of the function $f(z)$.
In this parametrization the black hole temperature and $U(1)$ chemical potential --- equal to the CFT temperature and bare chemical potential --- are
\begin{eqnarray}
  \label{eq:19}
  T_{CFT}=T_{BH}= \frac{\alp}{4\pi}(3-q^2)~,~~
  \mu_{0}=\mu_{BH}= -2q\alp~.
\end{eqnarray}
The parameter $q$ is bounded between $0\leq q^2\leq 3$
interpolating between AdS-Schwarzschild and the extremal AdS black
hole. 
For the equation of motion of fermions in this background we
shall need the spin connection belonging to this metric. The nonzero
components are
\begin{eqnarray}
  \label{eq:2}
  \ome_0^{ab} = 
%\delta_0^{[a}\del_{z}^{b]} \frac{z\sqrt{f}}{L}
%  \pa_z( \frac{L\alp}{z} \sqrt{f})  &,&
%\non
% &=& 
- \delta_0^{[a}\del_{z}^{b]} \alp f\left(\frac{1}{z}-\frac{\pa_zf}{2f}\right)
%\non
&,&
  \ome_i^{ab} = 
%\delta_i^{[a}\del_{z}^{b]} \frac{z\sqrt{f}}{L}
%  \pa_z( \alp\frac{L}{z}) \non 
%&=& 
- \delta_i^{[a}\del_{z}^{b]} \frac{\alp\sqrt{f}}{z}~.
\end{eqnarray}
%Note the fact that the warp function $f$ appears in fractional powers
%$\sqrt{f}$. As we will see, this will cause the field equation for the
%spinors to have peculiar behaviour...

%\def\slT{\,\slash\!\!\!\!T\,}
\def\slT{\,\slash\!\!\!\! \cT}
\def\slD{\slash\!\!\!\!D}

Applying the AdS${}_{d+1}$/CFT${}_d$ dictionary the CFT fermion-fermion correlation function is computed from the action $S_{bulk}$ (\ref{eq:3}) supplemented by appropriate boundary terms, $S_{bdy}$. One constructs the on-shell action given an arbitrary set of fermionic boundary conditions and the latter are then interpreted as sources of fermionic operators in the CFT:
\begin{eqnarray}
  \label{eq:4}
 Z_{CFT}(J)= \langle e^{J\cO} \rangle_{CFT} &=& \left. \exp\left[i(S_{bulk}+S_{bdy})^{on-shell}(\phi(J))\right]\right|_{\left.\phi\right|_{\pa AdS} = J} ~.
\end{eqnarray}
The issue of which boundary terms ought to be added to the bulk action tends to be subtle. For fermionic systems it is critical as the bulk action (\ref{eq:3}) identically vanishes on-shell \cite{Henningson:1998cd,Mueck:1998iz,Henneaux:1998ch,Contino:2004vy,Kirsch:2006he}. Because the field equations for the fermions are first order and half the components of the spinor correspond to the conjugate momenta of the other half, we can in fact only choose a boundary source for half the components
of $\Psi$. Projecting onto eigenstates of $\Gam^z$,
$\Gam^z\Psi_{\pm}=\pm\Psi_{\pm}$, we will 
choose a boundary source $\Psi_+^0\equiv
\lim_{z_0\rar 0} \Psi_+(z=z_0)$  (to regulate the theory we impose the boundary conditions at a small distance $z_0$ away from the formal boundary $z=0$ and take $z_0\rar 0$ at the end). The boundary value $\Psi_-^0$ is not independent but related to that of $\Psi_+^0$ by the Dirac equation. We should therefore not include it as an independent degree of freedom when taking functional derivatives with respect to the source. Adding a boundary action,
\begin{eqnarray}
  \label{eq:26}
  S_{bdy} &=& \frac{L^2}{2\kap_4^2}\int_{z=z_0}\!\!\!\!\!d^3x\, \sqrt{-h}\,\bar{\Psi}_+\Psi_-
\end{eqnarray}
with $h_{\mu\nu}$ the induced metric
ensures a proper variational principle \cite{Contino:2004vy}. The variation of $\del\Psi_-$ from the boundary action,
\begin{eqnarray}
  \label{eq:28}
  \del S_{bdy} &=& 
%\frac{L^3\alp^3}{z^3} \int_{z=z_0}d^3x\,\left[ \del\bar{\Psi}_+\Psi_- + %\bar{\Psi}_+\del\Psi_-\right]  \non 
 \left.\frac{L^2}{2\kap_4^2}\int_{z=z_0}\!\!\!\!\!d^3x \sqrt{-h}\,\bar{\Psi}_+\del\Psi_-\right|_{\Psi_+^0 {\rm fixed}}~,
\end{eqnarray}
now cancels the boundary term from variation of the bulk action
\begin{eqnarray}
  \label{eq:27}
  \del S_{bulk} &=& \frac{L^2}{2\kap_4^2} \int \sqrt{-g}\left[-\del\bar{\Psi}(\slD+m)\Psi_+ -\overline{\left((\slD+m){\Psi}\right)}\del\Psi\right] \non
&&+  \frac{L^2}{2\kap_4^2}\left.\int_{z=z_0} \sqrt{-h}\left[ -\bar{\Psi}_+\del\Psi_--\bar{\Psi}_-\del\Psi_+  \right]\right|_{\Psi_+^0 {\rm fixed}}~.
\end{eqnarray}

\subsection{The Fermion Green's function.}
To compute the fermion Green's function, we thus solve the field-equation for $\Psi(z)$ with  $\Psi_+^0\equiv
\lim_{z_0\rar 0} \Psi_+(z=z_0)$  as the boundary condition, substitute this solution back into the combined action $S_{bulk}+S_{bdy}$ and functionally differentiate twice.
As $\Psi_{sol}(\Psi^0_+)$ obeys the field-equation --- the Dirac Equation ---
\begin{eqnarray}
  \label{eq:22}
  (\slD+m)\Psi^{sol}(\Psi_+^0)=0~,
\end{eqnarray}
the contribution to the on-shell action is solely due to the boundary action $S_{bdy}$ in eq. (\ref{eq:26}).
To solve the Dirac equation, we Fourier transform along the boundary, 
\begin{eqnarray}
\Psi(z,x^i,t)=\int \frac{d\ome d^2k}{(2\pi)^3}\,\,\Psi(z,k_i,\ome) e^{ik_ix^i-i\ome t}~,
\end{eqnarray}
 and
project onto the eigenstates of $\Gam^z$,
$\Gam^z\Psi_{\pm}=\pm\Psi_{\pm}$. 
Choosing the basis of Dirac matrices
\begin{eqnarray}
&&\Gam^z=\sig^3\otimes\one ~,~~
\Gam^i=\sig^1\otimes \sig^i~,~~
\Gam^t=\sig^1\otimes \sig^t =\sig^1\otimes i\sig^3 ~,
\end{eqnarray}
we can consider $\Psi_{\pm}$ to be two-component Dirac spinors
appropriate for $d=3$ from here on. In the charged AdS black hole background  with non-zero gauge field from eq. (\ref{eq:1}), the Dirac equation decomposes into the two equations
\begin{eqnarray}
  \label{eq:10}
  (\pa_z+\cA^{\pm})\Psi_{\pm} = \mp\slT\Psi_{\mp}~,
\end{eqnarray}
with
\begin{eqnarray}
  \label{eq:11}
  \cA^{\pm} &=& -\frac{1}{2z}\left(3-\frac{z\pa_z f}{2f}\right)\pm
  \frac{Lm}{z\sqrt{f}} ~,\non
  \slT &=& \frac{i}{\alp
    f}\left[\left(-\ome+2gq\alp (z-1)\right)\sig^t+\sqrt{f}k_i\sig^i\right]~.
\end{eqnarray}
We can eliminate either $\Psi_+$ or $\Psi_-$ and readily derive
a second order dynamical equation for $\Psi_{\pm}$. Using that
\begin{equation}
\label{eq:16}
\slT\slT=-T_tT_t+T_1T_1+T_2T_2\equiv T^2~,
\end{equation} 
we can invert $\slT$ to rewrite
\begin{eqnarray}
  \label{eq:12}
  \frac{\slT}{T^2}\left(\pa_z+\cA^+\right)\Psi_+ = -\Psi_-
\end{eqnarray}
and use the $\Psi_-$ equation to obtain 
\begin{eqnarray}
  \label{eq:13}
  (\pa_z+\cA^-)\frac{\slT}{T^2}(\pa_z+\cA^+)\Psi_+ = -\slT\Psi_+~.
\end{eqnarray}
Using the identity eq. (\ref{eq:16}) repeatedly, this is equivalent to
\begin{eqnarray}
  \label{eq:14}
\left(\pa_z^2+P(z)\pa_z +Q(z)\right)\Psi_+=0
\end{eqnarray}
with
\begin{eqnarray}
\label{eq:15}
&&P(z) = (\cA^-+\cA^+)-[\pa_z,\slT]\frac{\slT}{T^2}~, \non
&&Q(z) = \cA^-\cA^++(\pa_z\cA^+)-[\pa_z,\slT]\frac{\slT}{T^2}\cA^++T^2~.
\end{eqnarray}
Note that both $P(z)$ and $Q(z)$ are two-by-two matrices.
The equation for $\Psi_-$ is simply obtained by switching $\cA^+$ with
$\cA^-$ and $\slT$ with $-\slT$; it is the CPT conjugate obtained by sending $m \rar -m$ and $\{\ome,k_i,q\}\rar \{-\ome,-k_i,-q\}$.

\newcommand{\upone}{\tiny \matrix{1\cr 0}}
\newcommand{\downone}{\tiny \matrix{0\cr 1}}

We can now derive a formal expression for the propagator in terms of the solutions to the second-order equation. We write the on-shell bulk field as
\begin{eqnarray}
  \label{eq:7}
  \Psi_{+}^{sol}(z) &=& F_{+}(z)F^{-1}_{+}(z_0)\Psi^0_{+}(z_0) ~
\end{eqnarray}
where $F_{\pm}(u)$ is the two-by-two matrix satisfying the second
order equation (\ref{eq:14}) \cite{Contino:2004vy} subject to a boundary condition in the interior of AdS. We will discuss the appropriate interior boundary condition below. There are two independent solutions $\Psi^{\left(\upone\right)}_+(z),~\Psi^{\left(\downone\right)}_+(z)$ that obey the interior boundary condition, one for each component of the spinor. In terms of this solution the matrix $F_+(z)$ equals 
\begin{eqnarray}
F_+(z) =\left(\matrix{\Psi^{\left(\upone\right)}_+(z)~,~~ \Psi^{\left(\downone\right)}_+(z)}\right)~.
\end{eqnarray}
%\comment{MOVE THIS TO LATER: MASSES AND DIMENSIONS}
Similarly for $\Psi_-^{sol}(z)$ we write 
\begin{eqnarray}
  \label{eq:23}
  \Psi_-^{sol}(z)=F_-(z)F^{-1}_-(z_0)\Psi^0_-(z_0)~.
\end{eqnarray}
However, $\Psi_-^0$ is not independent as we emphasized earlier. It is related to $\Psi_+^0$ through the Dirac equation in its projected form
(\ref{eq:10}). Acting with $\pa_z+A^+$ on both sides of (\ref{eq:7})
we see that \cite{Contino:2004vy}
\begin{eqnarray}
  \label{eq:8}
  &(\pa_z+A^+)\Psi_+^{sol} &=
  (\pa_z+A^+)F_{+}(z)F^{-1}_{+}(z_0)\Psi^0_+ \non
\Leftrightarrow&~~~
  -\slT\Psi_-^{sol}&= -\slT F_{-}(z)F^{-1}_+(z_0)\Psi_+^0~.
\end{eqnarray}
We have used that all the $z$ dependence of $\Psi^{sol}_{\pm}(z)$ is encoded  in the matrices $F_{\pm}(z)$ and therefore $F_{\pm}(z)$ obey the same projected Dirac equations. Thus we find that $\Psi^0_-$ equals
\begin{eqnarray}
\Psi_-^0 &=& F_-(z_0) F^{-1}_+(z_0)\Psi_+^0~.
\end{eqnarray}
Substituting this constraint into the boundary action, we obtain an expression for the full on-shell action in terms of the solutions $F_{\pm}(z)$:
\begin{eqnarray}
  \label{eq:6}
  S^{on-shell} = \frac{L^2}{2\kap_4^2} \int_{z=z_0}\frac{d\ome d^2k}{(2\pi)^3} \sqrt{-h} \,\, \bar{\Psi}^0_+\, F_-(z_0)F_+^{-1}(z_0)  \,\Psi^0_+~.
\end{eqnarray}
Up to a normalization $\cN$ the two-point function is therefore
\begin{eqnarray}
  \label{eq:9}
  G(\ome,k) = \frac{1}{\cN} F_-(z_0)F_+^{-1}(z_0)~.
\end{eqnarray}
This is the time-ordered two-point function. For the spectral function we shall need the imaginary part of the retarded propagator. At finite temperature the AdS background is no longer regular in the interior but has a horizon. In principle one should also consider its boundary contribution. The retarded propagator prescription of \cite{Son:2002sd} --- verified in \cite{Herzog:2002pc}--- is to ignore this contribution and to impose infalling boundary conditions at the horizon instead of regularity at the center of AdS. 
This is what we shall do.

The retarded Green's function for fermions is still a matrix. Parity and rotational invariance dictate that it can be decomposed as
\begin{eqnarray}
  \label{eq:18}
  G_R(\ome,k) = \Pi_s + \sig^t\Pi_t+\sig^i\Pi_i~.
\end{eqnarray}
Our main interest, the spectral function, proportional to Im$\langle \Psi^\dagger\Psi \rangle$, is the imaginary part of $\Pi_t$. Specifically
\begin{eqnarray}
  \label{eq:24}
  A(\ome,k) = -\frac{1}{\pi}{\rm Im} ({\rm Tr} \,i\sig^t G_R(\ome,k))~.
\end{eqnarray}
%\comment{MOVE to SoM}
As a consequence of the underlying conformal symmetry both the Green's function and the spectral function possess a scaling symmetry. Eq. (\ref{eq:11}) shows that the frequency $\ome$ and momenta $k$ are naturally expressed in units of an effective temperature $T_{eff}(\mu_0)\equiv 3 \alp/4\pi $ which depends on the chemical potential $\mu_0$
$$
T_{eff}(\mu_0)=T\left(\frac{1}{2}+\frac{1}{2}\sqrt{1+\frac{(\mu_0\sqrt{3})^2}{(4\pi T)^2}}\right)~.
$$
The spectral function computed from AdS is therefore naturally of the form
%\footnote{We have rescaled $\mu \rar 4\pi \mu/\sqrt{3}$, $\ome \rar 4\pi \ome/3$, $k \rar 4\pi k/3$ with respect to the natural AdS units given in the supporting material.}
$$
%  \label{eq:33}
  A_{\Delta_{\Psi}}^{\frac{\mu_0}{T}}(\ome,k) =\frac{1}{T^{d-2\Delta_{\Psi}}} \tilde{f} \left(\frac{\ome}{T_{eff}(\mu_0)},\frac{k}{T_{eff}(\mu_0)}; \Delta_{\Psi}, \frac{\mu_0}{T}\right)~.
$$
Any rescaling of $T_{eff}$ can be compensated by a rescaling of the frequencies and momenta and $\mu_0/T$ is the single independent parameter determining the characteristics of the fermion spectral function. The results in the main text have been converted to units of $k/T$ or $k/\mu_0$ for clarity of the presentation.

\subsection{Masses and Dimensions.}

A final crucial step is the establish the aforementioned relation between the mass of the AdS fermion and the scaling dimension of the dual fermionic operator in the CFT. For generality we shall work in $d$ dimensions in this subsection. This subsection recapitulates \cite{Contino:2004vy}.

The scaling behavior can be read off from the asymptotic behavior of the solution near the boundary $z=0$. In this limit the second order equation (\ref{eq:14}) diagonalizes: (setting $L=1$)
\begin{eqnarray}
  \label{eq:21}
\left(
  \pa_z^2-\frac{d}{z}\pa_z+\frac{d(d+2)-4m(1+m)}{4z^2}\right)\Psi_+ = 0+\ldots
\end{eqnarray}
Clearly the temperature or chemical potential of the black-hole is immaterial to the asymptotic scaling behaviour at $z=0$; in terms of the CFT $z=0$ is the UV of the theory and it should be insensitive to the infrared physics at the horizon. The leading powers of the two independent solutions to this equation are
\begin{eqnarray}
  \label{eq:25}
  \Psi_+ (z)= z^{\frac{d+1}{2}-|m+\frac{1}{2}|}(\psi_++\ldots)+z^{\frac{d+1}{2} +|m+\frac{1}{2}|}(A_{+}+\ldots)~.
\end{eqnarray}
(we may drop the absolute value signs in principle, but as it emphasizes the special value $m=-1/2$ it will be instructive to keep them.)
Similarly for $\Psi_-(z)$ the leading singularities are obtained by sending $m\rar -m$
\begin{eqnarray}
  \label{eq:25}
  \Psi_- (z)= z^{\frac{d+1}{2}-|m-\frac{1}{2}|}(\psi_-+\ldots)+z^{\frac{d+1}{2} +|m-\frac{1}{2}|}(A_{-}+\ldots)
\end{eqnarray}
However, recall that the Dirac equation relates the two asymptotic behaviors and that the boundary value of $\Psi_-$ is not independent. Near $z=0$
\begin{eqnarray}
  \label{eq:29}
  (\pa_z -\frac{d/2-m}{z}) \Psi_+ = -\slT|_{z=0} \Psi_-+\ldots
\end{eqnarray}
Thus $\psi_- \propto \psi_+$ and $A_-\propto A_+$.

Because the equation diagonalizes, each component of $\Psi_\pm(z)$ can be considered independently and the matrices $F_{\pm}(z)$ diagonalize in the limit $z\rar 0$. The scaling behavior of the Green's function is then readily read off from its definition 
\begin{eqnarray}
  \label{eq:30}
  G(\ome,k) = \frac{1}{\cN}F_-F_+^{-1}~ \sim~ \frac{z^{\frac{d+1}{2}-|m-\frac{1}{2}|}(\psi_-+\ldots)+z^{\frac{d+1}{2} +|m-\frac{1}{2}|}(A_{-}+\ldots)}{z^{\frac{d+1}{2}-|m+\frac{1}{2}|}(\psi_++\ldots)+z^{\frac{d+1}{2} +|m+\frac{1}{2}|}(A_{+}+\ldots)} ~.
\end{eqnarray}
The dominant scaling behavior depends on the value of $m$ and there are three different regimes $(I)$: $m>\frac{1}{2}$, $(II)$: $\hlf>m>-\hlf$, and $(III)$: $-\hlf>m$. In these regimes the Green's function behaves as
\begin{eqnarray}
  \label{eq:31}
  G(\ome,k) \sim 
\left\{
\matrix{ 
z\left(\frac{\psi_-}{\psi_+} + \ldots\right) 
+ z^{2m}\left(\frac{A_-}{\psi_+} +\ldots \right) ~~~~~& m>\hlf~, \cr
z^{2m}\left(\frac{\psi_-}{\psi_+}+\ldots\right)
+z\left(\frac{A_-}{\psi_+} +\ldots \right) ~~~~~~ &\hlf>m>-\hlf~, \cr
\frac{1}{z}\left(\frac{\psi_-}{\psi_+} + \ldots\right) 
+ \frac{1}{z^{2m}}\left(\frac{A_-}{\psi_+} +\ldots \right) & -\hlf>m~.} 
\right.
\end{eqnarray}
In regime (I) the contribution proportional to $z$ yields a contact term \cite{Contino:2004vy}. Recall that at zero-temperature and chemical potential each power of $z$ is accompanied by a power of momentum: the dimensionless arguments of the solutions $\Psi^{sol}_{\pm}(z)$ are $kz$ and $\ome z$. Discarding the term analytic in $z$ and thus analytic in momenta, the second term proportional to $z^{2m}$  yields a Green's function
\begin{eqnarray}
  \label{eq:32}
  G(\ome,k) \sim  (z_0\ome)^{2m}
\end{eqnarray}
corresponding to the two-point function of a conformal operator of weight $\Delta_{\Psi}=\frac{d}{2}+m$. In regime (II) there is no contact term and one immediately finds the same relation between the AdS fermion mass and the scaling dimension of the conformal operator. In regime (III), however, one finds an explicit pole $(\ome z)^{-1}$ independent of the AdS fermion mass or the spacetime dimension. It signals an inconsistency in the theory and one cannot consider this regime as physical \cite{Contino:2004vy}. This is reminiscent of the situation for scalars where for $m^2_{scalar}>-d^2/4+1$ one finds analytic terms in the two-point correlator; for $-d^2/4+1>m^2_{scalar}>-d^2/4$ both solutions are normalizable; and for $-d^2/4>m^2_{scalar}$ the theory is inconsistent.
The analogy with scalars may appear strange since a negative mass-squared for scalars clearly can be problematic, whereas the sign of the fermion-mass term does not have any physical consequences normally. Recall, however, that the same AdS bulk action can describe several CFTs depending on the boundary terms added to the action \cite{Klebanov:1999tb}. We have chosen a very specific boundary action such that $\Psi_+(z)$ is the independent variable which breaks the degeneracy between (bulk) theories with $m>0$ and $m<0$.
In this theory $m$ is bounded below by $-1/2$. We could have chosen a different theory with $\Psi_-(z)$ the independent variable. One would find then that $m$ is bounded from above by $1/2$. The regime $1/2>m>-1/2$ is present in both theories; it is the range where both solutions are normalizable and choosing either $\Psi_+(z)$ or $\Psi_-(z)$ as the independent variable corresponds to switching the ``sources'' and ``expectation values'' in the usual way (see also \cite{Iqbal:2009fd}).

This analysis also teaches us that the normalization $\cN$ should go as $z_0^{2m}$ to obtain a finite answer in the limit $z_0\rar 0$.\footnote{Note that the factor $L^2/2\kap_4^2$ in the on-shell action (\ref{eq:6}) follows from an unconventional normalization of the fields in the action (\ref{eq:3}). It would be absent for conventional normalization.}

% The choice which value to use for $m$ will prove essential to show the emergence of the Fermi liquid. The lower end of the unitarity bound $m=-1/2$
% correspond to introducing a fermionic conformal operator with weight  $\Delta=(d-1)/2$. This equals the scaling dimension of a free fermion. Despite the fact that the underlying CFT is strongly coupled, the absence of a large anomalous dimension for a fermion with mass $m=-1/2$ argues that such an operator fulfills a spectator-role and is only semi-weakly coupled to this CFT. {\em We will therefore use  this value throughout most of our calculations.} Our expectation is that the Fermi liquid, as a
% system with well-defined quasiparticle excitations, can be
% described in terms of weakly interacting fields. As we increase $m$ and we move away from the free-field value of the scaling dimension we expect this reasoning to value. At precisely $m=0$ we expect the Fermi-liquid to disappear. For this value the naive scaling dimension of the Fermion-bilinear is marginal and this argues that the vacuum of the theory remains is not affected. This intuition will be borne out by our results.  

\subsection{The retarded propagator boundary conditions at the horizon.}

The final component of our set-up will be the boundary conditions at the horizon of the the black hole. To compute the retarded propagator in thermal settings/black hole the appropriate b.c. are those
infalling into the horizon. Near the horizon at $z=1$, the second
order equation for $\Psi_{\pm}$ becomes the same for both $\Psi_+$ and $\Psi_-$ and moreover diagonalizes:
\def\tome{\tilde{\ome}}
\begin{eqnarray}
  \label{eq:17} \left(\pa_z^2-\frac{3}{2(1-z)}\pa_z+\frac{\tome^2+\frac{1}{16}}{(1-z)^2}\right)\Psi_{\pm}+\cO((z-1))=0.
\end{eqnarray}
with $\tome \equiv \frac{\ome}{a(3-q^2)} =\frac{\ome}{4\pi T}$. This equation has solutions of the form
\begin{eqnarray}
  \label{eq:5}
  \Psi_{\pm}= (1-z)^{i\tome-\frac{1}{4}}(c_r+...)+(1-z)^{-i\tome-\frac{1}{4}}(c_i+...)
\end{eqnarray}
The second solution has the incoming boundary condition we seek.\footnote{
A technical detail is that due to the factors $\sqrt{f}$ in the field equation, there is
no standard Frobenius solution $\Psi_{\pm} =  (1-z)^{\pm
  i\tome-\frac{1}{4}}\sum_{n=0}^{\infty}a_n^{(\pm)}(1-z)^n$. Rather
half-integer powers of $(1-z)$ appear as well. We need the
Frobenius method for the numerics: we use it to construct a second b.c. for the derivative of $\Psi_+$ --- see e.g. \cite{CaronHuot:2006te}. Changing coordinates to $z=1-s^2$ solves this problem.}

\subsection{ A comment on the special case $\Delta_{\Psi}=3/2$.}

In the specific case of zero AdS Dirac mass, or $\Delta_{\Psi}=3/2$ the spectral function also shows a distinctive delta-function-like peak studied in \cite{Liu:2009dm}. However, this is not the Fermi-liquid quasiparticle --- the quasiparticle pole strength vanishes exponentially as $\Delta_{\Psi} \rar 3/2$. Already for $\Delta_{\Psi} < 3/2$ this other peak can be identified in the spectral function in the frequency domain for fixed $k$, although it has negligible weight compared to the quasiparticle (See supporting Figure S.1). It is quite a mystery what this peak signifies in the regime $\Delta_{\Psi} < 3/2$. It does not seem to be another quasiparticle peak; it is always located precisely at $\ome=0$ for any $k$ --- it does not disperse and it  always has finite positive energy relative to the true zero energy quasiparticle located at $E_F<0$. 
The lack of dispersion indicates that this peak corresponds to a state that is rather localized in real space while evidently breaking conformal invariance, but a true understanding of this mysterious state is an open question.

\begin{figure}
\centering
\includegraphics[width=4.75in]{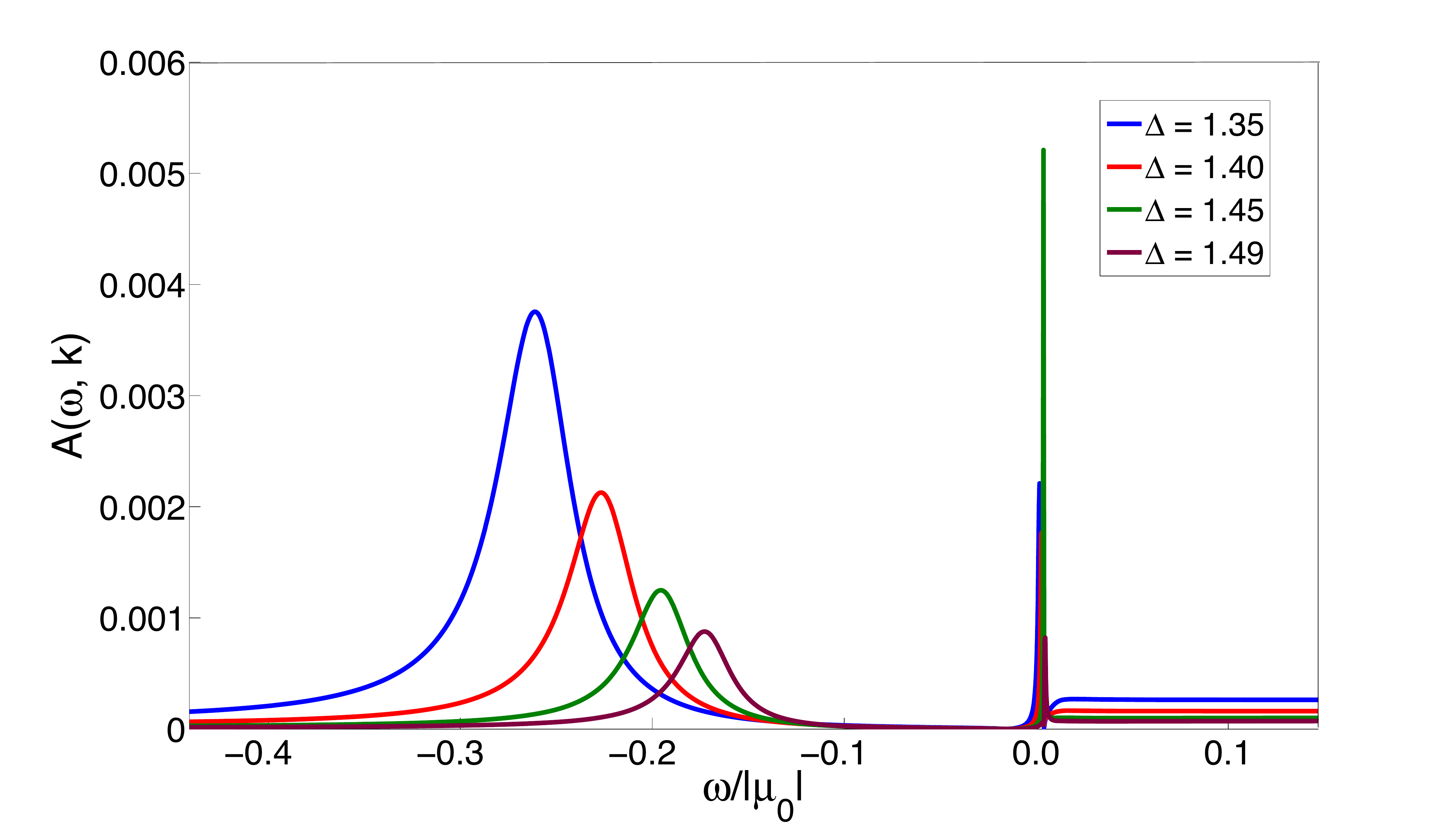}
\caption{\small \it 
{\em Supporting Figure}
The relation between the quasiparticle peak and the $\ome=0$ localized peak as $\Delta_{\Psi} \rar 3/2$ for $k/\mu_0=0.35$ and $\mu_0/T=-30.9$.
}
\end{figure}

% \begin{scilastnote}
% \item
{\bf Acknowledgements:} We thank F. Denef, S. Hartnoll,
 H. Liu, J. McGreevy, S. Sachdev, D. Sadri and D. Vegh for discussions. This research was
supported in part by a VIDI Innovative Research Incentive Grant
(K. Schalm) from the Netherlands Organisation for Scientific
Research (NWO), a Spinoza Award (J. Zaanen) from the Netherlands
Organisation for Scientific Research (NWO) and the Dutch
Foundation for Fundamental Research on Matter (FOM).
%\end{scilastnote}
%\begin{thebibliography}{999}
%%%%%%%%%%%%%%%%%% The ones below are from the supporting material

% \end{thebibliography}

% \end{document}

%\noindent 
{\small
%\baselineskip12pt

% \begin{scilastnote}
% \item
% {\bf Acknowledgements:} We thank F. Denef, S. Hartnoll,
%  H. Liu, J. McGreevy, S. Sachdev, D. Sadri and D. Vegh for discussions. This research was
% supported in part by a VIDI Innovative Research Incentive Grant
% (K. Schalm) from the Netherlands Organisation for Scientific
% Research (NWO), a Spinoza Award (J. Zaanen) from the Netherlands
% Organisation for Scientific Research (NWO) and the Dutch
% Foundation for Fundamental Research on Matter (FOM).
% \end{scilastnote}
}


\begin{thebibliography}{999}
\bibitem{footnote1}
ARPES Fermi-surface measurements assume that electrons are the only relevant charged objects. If this is so, then it measures the electron i.e. fermion spectral function. This spectral function is what we compute here, even though in our AdS set-up the fermions are almost certainly not the only charged objects.
\bibitem{footnote2}
Note that the $m=0$ spectral peak discussed in the article \cite{Liu:2009dm} is therefore not the peak we identified with the quasiparticle state. See supporting material~\cite{supp}.
\end{thebibliography}
\end{document}